\documentclass[11pt,aps,prc,amsmath,superscriptaddress,nofootinbib,tightenlines]{revtex4-2}

\usepackage[group-digits = false]{siunitx}
\usepackage{verbatim}
\usepackage{physics}
\usepackage[version=4]{mhchem}

\usepackage{amsmath}
\usepackage{amssymb}
\usepackage{mathtools}
\usepackage{bbold}

\usepackage{booktabs}
\usepackage{tabularx}
\usepackage{float}

\usepackage{ifthen}

\usepackage{hyperref}
\hypersetup{colorlinks=true, citecolor=blue, urlcolor=blue, linkcolor=blue}
\usepackage{cleveref}

\usepackage{orcidlink}

\usepackage{tikz}
\usetikzlibrary{arrows, decorations.pathmorphing}
\usetikzlibrary{shapes}
\newcommand{\midarr}{\tikz \filldraw (-.1,-.1) -- (.1,0) -- (-.1,.1) -- (-.1,-.1);}

\newcommand{\eps}{\varepsilon}

\renewcommand{\i}{\mathrm{i}}
\newcommand{\const}{\mathrm{const.}}
\renewcommand{\d}{\mathrm{d}}
\DeclareMathOperator{\arcosh}{\mathrm{arcosh}}

\let\Im\relax
\DeclareMathOperator{\Im}{\mathrm{Im}}

\newcommand{\E}[1]{\langle#1\rangle}

\renewcommand{\vec}[1]{\boldsymbol{#1}}

\newcommand{\mprescript}[3]{{{\vphantom{#3}}}^{#1}_{#2}\! #3 }

\newcommand{\ibraket}[4]{ \mprescript{}{#1}{ \braket{#2}{#3}}_{#4}^{}  } 
\newcommand{\ibra}[2]{ \mprescript{}{#1}{ \bra{#2} } }
\newcommand{\iket}[2]{ \ket{#1}_{#2}^{} } 
\newcommand{\iketbra}[3]{
  \ifthenelse{\equal{#3}{}}{
    \iket{#1}{#2} \ibra{#2}{#1}
  }
  {
    \iket{#1}{#2} \ibra{#2}{#3}
  }
}

\newcommand{\iketbrae}[3]{
  \ifthenelse{\equal{#3}{}}{
    \iket{#1}{#2} \ibra{#2}{#1}
  }
  {
    \iket{#1 \vphantom{#3}}{#2} \ibra{#2}{#3 \vphantom{#1}}
  }
}

\newcommand{\K}[1]{\left( #1 \right)}

\begin{document}

\title{Renormalizing Two-Neutron Halo Nuclei Without Neutron-Core Interaction}

\author{Daniel Kromm\,\orcidlink{0009-0009-2922-6314}}
\email[E-mail:]{daniel.kromm@tu-darmstadt.de}
\affiliation{Technische Universität Darmstadt, Department of Physics, Institut für Kernphysik, 64289 Darmstadt, Germany}

\author{Matthias Göbel\,\orcidlink{0000-0002-7232-0033}}
\email{goebel.matthias@ujf.cas.cz}
\affiliation{Nuclear Physics Institute, Czech Academy of Sciences, 25068  Řež, Czech Republic}
\affiliation{Istituto Nazionale di Fisica Nucleare, Sezione di Pisa, Largo Pontecorvo 3, 56127 Pisa, Italy}

\author{Hans-Werner Hammer\,\orcidlink{0000-0002-2318-0644}}
\email{Hans-Werner.Hammer@physik.tu-darmstadt.de}
\affiliation{Technische Universität Darmstadt, Department of Physics, Institut für Kernphysik, 64289 Darmstadt, Germany}
\affiliation{Helmholtz Forschungsakademie Hessen f\"ur FAIR (HFHF)
    and ExtreMe Matter Institute EMMI, GSI Helmholtzzentrum f\"{u}r Schwerionenforschung GmbH, 64291 Darmstadt, Germany}

\date{\today}

\begin{abstract}
    We consider the Effective Field Theory (EFT) {\color{black} scheme} proposed by Hongo and Son (HS) to describe two-neutron halo nuclei where the neutron-core interaction is subleading. In this EFT, the ratio of the mean-square matter radius and charge radius is universal in so far that it only depends on the two-neutron separation energy of the nucleus and the neutron-neutron scattering length.
    By investigating the divergence structure of this theory, we find that one further renormalization condition is required to predict both radii separately. Our renormalization scheme uses one of the mean square radii or the scattering amplitude as input. 
    We use the {\color{black} HS scheme} to calculate the matter radii of the two-neutron halo nuclei \(^{11}\)Li, \(^{14}\)Be, \(^{17}\)B, \(^{19}\)B, and \(^{22}\)C and compare to the values
    obtained with standard Halo EFT.
    In this comparison we use both the physical value of the neutron-core scattering length and rescaled values.
    We observe good convergence against the {\color{black} HS scheme} for the case of a negligible neutron-core interaction.
    Similar agreement for the radii is also found in the case of the halo nucleus \(^6\)He, where the \(nc\) interaction is in the p-wave.
    Our renormalization scheme makes the restriction in the ultraviolet cutoff range from the Landau pole explicit.
    We calculate the position of the Landau pole for various halo nuclei.
    In all cases the Landau pole restricts the cutoff to rather low values.
    Finally, we derive an explicit expression for the three-to-three neutron-neutron-core scattering amplitude and discuss its cut structure.
\end{abstract}


\maketitle


\section{Introduction}
\label{sec:intro}

Halo nuclei are exotic nuclear systems living close to the nuclear drip lines.
They are characterized by a "halo" of a few weakly bound halo nucleons and a more tightly bound core \cite{Tanihata:1985psr,Hansen:1987mc,Hansen:1995pu,Jonson:2004,Jensen:2004zz,Riisager:2012it,Hammer:2017tjm}. Here we focus on neutron halos.
As the halo neutrons spend most of their time in the classically forbidden region outside of the range of the interaction, halo nuclei live in the extreme quantum regime and show universal behavior \cite{Jensen:2004zz,braaten_dimer,Hammer:2017tjm}. They are typically short-lived, but can be produced and studied in-flight in rare-isotope beam factories \cite{Fahlander_2013}.
Typical examples of halo nuclei are \(^{11}\)Li and \(^{6}\)He.
Theoretically, halo nuclei have been successfully studied by employing cluster models.
Another powerful tool is Halo Effective Field Theory (Halo EFT), an EFT with cluster degrees of freedom. It can be seen as a systematization of cluster models with a well-defined power counting scheme. For a two-neutron halo nucleus, the effective degrees of freedom are the core and two neutrons such that the halo nucleus becomes an effective three-body system. The effective field theory treatment offers a systematic expansion of observables in the ratio of the low-energy scale over the high-energy scale.
The low-energy scale is associated with the typical halo dynamics, e.g., given by the two-neutron separation energy.
The counterpart is the high-energy scale defined as the lowest scale of omitted physics.  A typical example is the core excitation energy.
Effects of these omitted high-energy processes enter the description of low-energy dynamics via the so-called low-energy constants.
Halo EFT was first introduced for the \(^5\)He resonance \cite{Bertulani:2002sz,Bedaque:2003wa}.
In the last two decades it has been successfully applied to a number of different halo nuclei, among them two-neutron halo nuclei such as \(^{11}\)Li and \(^6\)He (see, e.g., Ref. \cite{Hammer:2017tjm} for a review).
Halo EFT has been applied to calculate a variety of observables such as matter and charge radii, \(E1\) strength distributions, \(\beta\)-decay lifetimes, and momentum distributions (see, e.g., Refs. \cite{Canham:2008jd,Hammer:2011ye,Gobel:2022pvz,Elkamhawy:2019nxq,Gobel:2021pvw}).

In standard Halo EFT the description of two-neutron halo nuclei is often done in terms of Faddeev amplitudes, which are determined by the Faddeev  equations.
Interactions enter the equations in terms of t-matrices.
The Faddeev amplitudes have to be calculated numerically and on the basis of those observables can be calculated.
Recently, Hongo and Son (HS) have constructed a special variant of Halo EFT for those two-neutron halo nuclei, in which the neutron-core interaction is subleading \cite{HongoSon22} (for further discussion of this EFT, see Refs.~\cite{naidon,Costa:2025ldi}).
{\color{black} This variant is based on the same degrees of freedon and symmetries as standard Halo EFT \cite{Bertulani:2002sz,Bedaque:2003wa,Hammer:2017tjm} but applies a different power counting regarding the neutron-core interaction. In this manuscript, we refer to their proposal as the "Halo EFT scheme of HS" or "HS scheme".}
The only remaining interactions at leading order are the neutron-neutron interaction and the three-body neutron-neutron-core interaction.
One advantage of this variant is that observables can be calculated largely analytically.
This also simplifies the study of correlations between different observables.
HS derived a relation for the ratio of root-mean-square (RMS) charge and matter radius. Moreover, they predicted the shape of the \(E1\) strength distribution, which parameterizes the Coulomb dissociation cross section.
However, calculating the charge and matter radii or the overall value of the \(E1\) strength directly is not possible in their
approach because of unknown couplings\footnote{
HS estimated the value of the unknown coupling to calculate the radii of \(^{11}\)Li.
}.

This paper investigates this issue in detail. We study alternative renormalization procedures for {\color{black} the HS Halo EFT scheme} to fully determine all couplings.
By analyzing the divergence structure in the cutoff, we show that another renormalization condition is necessary.
This analysis is supplemented by examples showing that two renormalization conditions alone do not lead to consistent predictions for the radii itself. Despite this, the ratio of charge and matter radii can be obtained even without full renormalization \cite{HongoSon22}.
We show that once we use either the charge or the matter radius as renormalization condition, we can determine the other radius for different two-neutron halo nuclei. Moreover, we calculate the radii for different halo nuclei, compare with radii obtained in standard Halo EFT, and investigate the constraints from the Landau pole.
Finally, we calculate the three-body scattering amplitude.

The paper is structured as follows.
We start with a review of  {\color{black} the HS scheme} and the original renormalization in Sec.~\ref{sec:review}.
Section~\ref{sec:reno} continues by discussing the possibility of alternative renormalization schemes.
We show that for full renormalization another renormalization condition is necessary.
Next, in Sec.~\ref{sec:full_reno}, we finalize the renormalization and illustrate the emergence of the Landau pole. 
Moving on to Sec.~\ref{sec:comparison}, we investigate the relation between charge radii and matter radii and compare to standard Halo EFT.
Finally, we calculate the three-body scattering amplitude in Sec.~\ref{sec:scattering}.
As a consistency check, we show that in the limit of a bound neutron-neutron system the unitary term has the correct form.
In Sec.~\ref{sec:concl}, we present our conclusions. Some technical and numerical details are discussed in the Appendices.

\section{Review of Theory by HS}\label{sec:review}

{\color{black}To keep our paper self-contained, we provide in this section a brief review of those aspects of {\color{black}the HS Halo EFT scheme} \cite{HongoSon22} which are relevant for our investigation.} We use units where $\hbar=c=1$.

\subsection{Two-Neutron System}
We start with the Lagrangian of the two-neutron system where we treat the neutrons as distinguishable particles labeled by their spins. {\color{black} This treatment of the neutron-neutron interaction is standard in short-range effective field theories, such as pionless EFT or Halo EFT. For more details, we refer the reader to Ref.~\cite{HongoSon22} and the recent reviews available on this subject \cite{Hammer:2017tjm,Hammer:2019poc}.}
It consists of a kinetic term for each neutron 
with mass $m_\mathrm{n}$ 
and a contact interaction term with a coupling $c_0$.
Since there is no mass renormalization in this theory, we can set $m_\mathrm{n}=1$ as the natural mass unit. {\color{black}In these units, the Lagrangian has dimension 5.} We use a Hubbard-Stratonovich transformation to introduce the dimer field $d$,
which is especially convenient because the only interaction of the core with neutrons contains all three particles. From now on we will use the term dimer to refer to the dineutron.
The Lagrangian becomes
\begin{equation}
\label{Lag_nn}
    \mathcal{L}_\mathrm{n}=\sum_{\alpha=\uparrow,\downarrow}\psi_\alpha^\dagger\left(\i\partial_t+\frac{\nabla^2}{2}\right)\psi_\alpha-\frac{1}{c_0}d^\dagger d+\psi_\uparrow^\dagger\psi_\downarrow^\dagger d+d^\dagger\psi_\downarrow\psi_\uparrow\,.
\end{equation}
From Eq.~(\ref{Lag_nn}) we see that the bare dimer propagator is constant, $-c_0$.
To obtain the dressed dimer propagator $D(p)$, we calculate the self-energy of the dimer.
The only diagram that contributes in this theory is the breakup of the dimer into two neutrons, which eventually recombine to a dimer. {\color{black}This dimer self-energy diagram is linearly divergent and must be regularized. 
We regularize the self-energy by introducing an ultraviolet (UV) momentum cutoff $\Lambda$. All terms proportional to inverse powers of the cutoff can be made arbitrarily small by choosing a sufficiently large cutoff. Thus we only need to keep the linear and constant terms in $\Lambda$. The dependence on $\Lambda$ can be removed by matching the dimer propagator 
to the t-matrix for s-wave neutron-neutron scattering at threshold, which is determined by the s-wave scattering length $a$. From this renormalization condition we get the running coupling $c_0=c_0(\Lambda)$. Next, we eliminate the coupling $c_0(\Lambda)$ and the cutoff-dependent terms in the self-energy in favor of $a$.
The final dressed dimer propagator, which is given by summing the geometric series of self-energy insertions to all orders, then reads \cite{HongoSon22}}
\begin{equation}
\label{eq:def_fa}
    D(p)=-4\pi f_a\left(\tfrac{\boldsymbol{p}^2}{4}-p_0-\i\eps\right),\quad \mbox{with}\quad f_a(x)=\left(\sqrt{x}-\frac{1}{a}\right)^{-1}\,.
\end{equation}

\subsection{Trimer Field}
To construct the EFT for the whole halo nucleus, we need to introduce further fields. Our Lagrangian is extended by a field $\phi$ describing the core and an auxiliary field $h$ describing the halo nucleus. The halo field $h$ is formally a dimer field as before but now composed of the core field $\phi$ and the neutron-neutron dimer field $d$. Since it essentially contains all three particles, we will refer to the halo field as the trimer field. Both new fields get a kinetic part with their respective masses $m_\phi=A$ and $m_h=A+2$. The field $h$ additionally receives a bare binding energy $B_0$. This is different from how we implemented the dimer field for the two neutrons. The derivatives of the halo field imply interactions of higher order between the core and the dimer. This would become apparent when transforming back into a pure dimer-core description without the trimer field.
The interaction between the respective fields of dimer and core is implemented via a contact interaction 
with a dimensionless coupling $g_0$. The resulting Lagrangian reads\footnote{Note that in their paper HS redefined the fields and the coupling using the field strength renormalization of the trimer field. We find it more transparent to postpone the introduction of the field strength renormalization to the point when we discuss observables.}
\begin{equation}\label{eq:lagrangian}
    \mathcal{L}=h^\dagger\left(\i\partial_t+\frac{\nabla^2}{2m_h}+B_0\right)h+\phi^\dagger\left(\i\partial_t+\frac{\nabla^2}{2m_\phi}\right)\phi+g_0(h^\dagger\phi d+\phi^\dagger d^\dagger h)+\mathcal{L}_\mathrm{n}\,,
\end{equation}
{\color{black}
where $m_\phi = A$ and $m_h = (A + 2)$ for core with mass number $A$. The dimensions of the fields are $[d] = 2$, $[\phi] = [\psi] = 3/2$. As a consequence, the 
$h\phi d$ coupling is dimensionless.
One may wonder why the Lagrangian for the trimer field contains a kinetic term at leading order. For the neutron-neutron dimer such a term appears only at higher order. The inclusion of the trimer kinetic term is mandated by the divergence structure of the trimer self-energy, which contains both a quadratic and a logarithmic divergence. The renormalization of the trimer self-energy and propagator will be discussed in detail in Secs. \ref{sec:reno} and \ref{sec:scattering}
below.
Apart from the specific implementation using a trimer field \cite{Bedaque:2002yg}, the main difference to standard Halo EFT  \cite{Bertulani:2002sz,Bedaque:2003wa,Hammer:2017tjm} is that the neutron-core interaction is suppressed and does not enter at leading order. In contrast, a three-body neutron-neutron-core interaction at leading order is standard in halo EFT. They have been considered previously for two-neutron halo nuclei with and without an alpha particle core \cite{Canham:2008jd,Rotureau:2012yu,Ji:2014wta,Ryberg:2017tpv}.}

\begin{figure}[t]
    \centering
    \begin{tikzpicture}
    \draw [double distance=1.5] (-1.3,0) -- (-1,0) node[pos=.2, left]{$h$};
    \draw [double distance=1.5] (1,0) -- (1.3,0) node[pos=.7, right]{$h$};
    \draw (1,0) arc (0:180:1) node[pos=.5]{\midarr} node[pos=.5, above]{$\phi$};
    \draw [dashed](-1,0) arc (-180:0:1) node[pos=.5]{\midarr} node[pos=.5, above]{$d$};
    \end{tikzpicture}
    \caption{\label{fig:trimer_self_energy}Self-energy diagram of the trimer field.}
\end{figure}
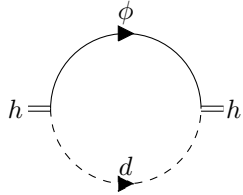

The self-energy $\Sigma_h(p)$ of the halo nucleus is given by one diagram, which is the breakup and recombination of the halo nucleus (see~Fig.~\ref{fig:trimer_self_energy}).
Since renormalization in this case is more delicate, we postpone its discussion and first perform calculations in terms of the bare parameters which are $B_0$ and $g_0$. The use of the dressed dimer propagator $D(p)$ ensures that all diagrams from the neutron sector $\mathcal{L}_\mathrm{n}$ are included. With this propagator and the Feynman rules derived from Eq.~\eqref{eq:lagrangian}, we get our initial expression for $\Sigma_h(p)$:
\begin{equation}\label{eq:halo_self_energy_first}
    \Sigma_h(p)=\i(\i g_0)^2\int\frac{\d^4q}{(2\pi)^4}\i G_\phi(q)\i D(p-q)\,.
    \end{equation}
Inserting both propagators, we evaluate the integral over $q_0$ via contour integration and arrive at a divergent three-momentum integral. Regularizing this integral with a sharp UV momentum cutoff $\Lambda$, we can write the self-energy as
\begin{equation}\label{eq:self-energy}
    \Sigma_h(p)=-4\pi g_0^2 I_\Lambda(-p_0+\tfrac{\boldsymbol{p}^2}{2m_h}-\i\eps)\,,
\end{equation}
where for a general argument\footnote{Note that $E$ appears as the argument of the integral and should not be confused with the energy of the trimer system. Especially $E=B>0$ does not correspond to a positive trimer energy.} $E\in\mathbb{C}\setminus(-\infty,0]$ the integral $I_\Lambda(E)$ is given by
\begin{equation}\label{eq:integral}
    I_\Lambda(E)=\int_{q\leq\Lambda}\frac{\d^3q}{(2\pi)^3}\frac{1}{\sqrt{E+\tfrac{\boldsymbol{q}^2}{2\mu}}-\frac{1}{a}}\,.
\end{equation}
This integral is quadratically divergent in the UV cutoff. We give a short derivation of this integral as well as closed form expressions in Appendix~\ref{ap:self-energy}. Summing the geometric series arising from the self-energy insertions, the full trimer Greens function is obtained as \cite{HongoSon22}:
\begin{equation}\label{eq:bare_trimer_prop}
    G_h^{-1}(p)=p_0-\frac{\boldsymbol{p}^2}{2m_h}+B_0+\i\eps+4\pi g_0^2I_\Lambda(-p_0+\tfrac{\boldsymbol{p}^2}{2m_h}-\i\eps)\,.
\end{equation}

\subsection{Radii and Universality}
Since $g_0$ and $B_0$ are the only parameters of the Lagrangian that are not fixed yet, we discuss how certain radii in this theory depend on them. The radii will of course naturally depend on our low-energy constants: the neutron-neutron scattering length $a$, the 2n separation energy $B$, and the core mass number $A$. For an unambiguous prediction, we need to replace any dependence on the first mentioned quantities by a dependence on $a$, $B$, and $A$ or other observables.
We start with the mean-square charge radius, which is determined by the spatial deviation of the charged core from the center of mass. Next, we do the same for the mean-square neutron radius given by the spatial extent of the dimer in order to calculate the mean-square matter radius of our halo nucleus. In either case the Feynman diagrams do not bring in further ambiguities regarding convergence of integrals. For more details regarding the calculation of radii and the geometry of the three-body system we refer to the supplementary material of Ref.~\cite{HongoSon22}.

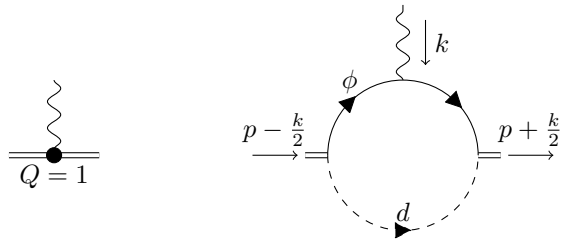
\begin{figure}[h]
	\centering
	$$
	\begin{tikzpicture}[baseline=-.0em]
	\draw [double distance=1.5] (-0.6,0) -- (0.6,0);
	\draw [decorate,decoration=snake] (0,1) -- (0,0);
	\draw [fill=black] (0,0) circle (0.1) node[below]{$Q=1$};
	\end{tikzpicture}
	\hspace{5em}
	\begin{tikzpicture}[baseline=-.0em]
		\draw [->] (-2, 0) -- (-1.4,0) node[pos=0.5, above]{$p-\tfrac{k}{2}$};
		\draw [->] (1.4, 0) -- (2,0) node[pos=0.5, above]{$p+\tfrac{k}{2}$};
		\draw [->] (.3,1.8) -- (.3,1.2) node[pos=0.5, right]{$k$};
		\draw [double distance=1.5] (-1.3,0) -- (-1,0);
		\draw [double distance=1.5] (1,0) -- (1.3,0);
		\draw (0,1) arc (90:180:1) node[pos=.5, rotate=45]{\midarr} node[pos=.5, above]{$\phi$};
		\draw [decorate,decoration=snake] (0,1) -- (0,2);
		\draw (1,0) arc (0:90:1) node[pos=.5, rotate=-45]{\midarr};
		\draw [dashed](-1,0) arc (-180:0:1) node[pos=.5]{\midarr} node[pos=.5, above]{$d$};
	\end{tikzpicture}
	$$
	\caption{Diagrams for the charge form factor of the halo nucleus. All external propagators are amputated. 
    For the detailed kinematics of the right-hand diagram see main text.}
	\label{fig:charge_form_factor}
\end{figure}

The charge radius is defined via the electric form factor of the halo nucleus, which can be computed from diagrams with one external photon that couples to the halo system. More precisely, one can extract the mean-square charge radius $\E{r_\mathrm{c}^2}$ from the electric form factor\footnote{The equation $F(\boldsymbol{k})=1-\frac{1}{6}\boldsymbol{k}^2\E{r_\mathrm{c}^2}+\mathcal{O}(k^4)$ stems from the Fourier transform of the charge density for low $\boldsymbol{k}$ in the case of a spherically symmetric charge density. But since we cannot assume our charge distribution to be spherically symmetric, we actually define $\E{r_\mathrm{c}^2}$ by this equation.} $F(\boldsymbol{k})=1-\frac{1}{6}\boldsymbol{k}^2\E{r_\mathrm{c}^2}+\mathcal{O}(k^4)$ for low photon momenta $|\boldsymbol{k}|$. To couple an external electromagnetic field $A_\mu$ to our system, one can apply the minimal coupling prescription $\i\partial_\mu\to\i\partial_\mu-\hat{Q}A_\mu$,
where $\hat{Q}$ is the charge operator\footnote{In principle, there can also be local gauge invariant terms not generated by minimal substitution. However, no such terms are present at this order.}.
The minimal substitution only applies to the core and halo field since these are in this EFT the only fields which both carry electric charge and appear with derivatives in our Lagrangian in Eq.~\eqref{eq:lagrangian}. There are two one-photon diagrams in our theory (see Fig.~\ref{fig:charge_form_factor}), which both are multiplied by the field strength renormalization $Z_h$ as both have external, amputated trimer propagators.

The first diagram is rather trivial, since the photon couples directly to the halo field. This diagram evaluates to $Z_h$. The second diagram contains the coupling to the core and therefore involves a one-loop calculation. This diagram is considered at the kinematic point
$k=(0,\boldsymbol{k})$ and $p=(-B+\tfrac{\boldsymbol{k}^2}{8m_h},\boldsymbol{0})$. This corresponds to the Breit frame, where the nucleus is reflected as the photon is absorbed. By expanding the form factor around $k=0$ one can check that the normalization condition $F(\boldsymbol{0})=1$ is fulfilled by the definition of the field strength renormalization $Z_h$ as the residue of the trimer propagator at the pole position
\begin{equation}\label{eq:field_strength_reno}
    Z_h=\left(\frac{\partial}{\partial p_0}G_h^{-1}(p_0=-B,\boldsymbol{p}=0)\right)^{-1}=\frac{1}{1-4\pi g_0^2I_\Lambda'(B)}\,.
\end{equation}
The mean-square charge radius itself can be extracted by expanding the form factor diagram up to order $k^3$. Comparing the terms quadratic in $k$ gives
\begin{equation}\label{eq:charge_radius}
    \E{r_\mathrm{c}^2}=Z_hg_0^2K_\mathrm{c}\,,
\end{equation}
where $K_\mathrm{c}$ is an explicit function only depending on $B$, $a$ and the core to neutron mass ratio $A$ \cite{HongoSon22}:
\begin{align}
    K_\mathrm{c}&=\frac{4A^{1/2}}{\pi(A+2)^{5/2}B}f_\mathrm{c}(\beta)
\end{align}
\begin{equation}
    f_\mathrm{c}(\beta)=\begin{cases}
\frac{1}{1-\beta^2}-\frac{\beta\arccos\beta}{(1-\beta^2)^{3/2}}\,,\quad&\beta<1\,,\\[5pt]
-\frac{1}{\beta^2-1}+\frac{\beta\arcosh\beta}{(\beta^2-1)^{3/2}}\,,\quad&\beta>1\,,
\end{cases}
\end{equation}
where $\beta:=-\tfrac{1}{a\sqrt{B}}$. The main result here is Eq.~\eqref{eq:charge_radius}. The charge radius is proportional to $Z_hg_0^2$. This statement also turns out to be true for the mean-square neutron radius $\E{r_\mathrm{n}^2}$. The neutron radius can similarly be defined by a form factor, which can be calculated from Feynman diagrams. But the external current now only couples to the neutrons in the dimer\footnote{One has to apply the minimal coupling with the charge operator replaced by an operator that counts the number of neutrons from the dimer. The neutron number for the core is zero.}. However, it certainly yields information about the spatial extent of the dineutron in the halo nucleus. The respective neutron form factor has the same expansion for low $|\boldsymbol{k}|$, but is normalized to $F_\mathrm{n}(\boldsymbol{0})=2$, since we have two neutrons. For the diagrams and the calculational details we again refer to Ref.~\cite{HongoSon22}. Since now the matter extent in our halo nuclei is determined by the spatial extent of dimer and core, the matter radius can be computed via an appropriately weighted average of the charge and neutron radii:
\begin{equation}
	\E{r_\mathrm{m}^2}=\frac{2}{A+2}\E{r_\mathrm{n}^2}+\frac{A}{A+2}\E{r_\mathrm{c}^2}=Z_hg_0^2K_\mathrm{m},
\end{equation}
where $K_\mathrm{m}$ is again an explicit function of $B$, $a$ and $A$ \cite{HongoSon22}:
\begin{align}
    K_\mathrm{m}=\frac{1}{\pi B}\left(\frac{A}{A+2}\right)^{3/2}\left(f_\mathrm{n}(\beta)+\frac{A}{A+2}f_\mathrm{c}(\beta)\right)
\end{align}
\begin{equation}
    f_\mathrm{n}(\beta)=\begin{cases}
\frac{1}{\beta^3}\left[\pi-2\beta+(\beta^2-2)\frac{\arccos\beta}{\sqrt{1-\beta^2}}\right],\quad&\beta<1\,,\\[5pt]
\frac{1}{\beta^3}\left[\pi-2\beta+(\beta^2-2)\frac{\arcosh\beta}{\sqrt{\beta^2-1}}\right],\quad&\beta>1\,.	
\end{cases}
\end{equation}
The mean-square matter radius is again proportional to $Z_hg_0^2$. Because of this factor, all our radii are not fully expressed so far in terms of low-energy constants. But dividing the radii, we get an expression independent of the factor $Z_hg_0^2$. Thus, we can state a well defined ratio of matter and charge radius \cite{HongoSon22}
\begin{equation}\label{eq:ratio_univ}
    \frac{\E{r_\mathrm{m}^2}}{\E{r_\mathrm{c}^2}}=\frac{K_\mathrm{m}}{K_\mathrm{c}}\,,
\end{equation}
which besides the mass number $A$ solely depends on $a$ and $B$. Hence, this ratio poses a universal prediction in the {\color{black} scheme of HS} \cite{HongoSon22}.

\section{Renormalization and Divergence Structure}\label{sec:reno}
In order to analyze the predictive power of the {\color{black} HS scheme} in detail, it is necessary to state renormalization conditions for the trimer. In particular, we want to find out how the variables $Z_h$ and $g_0$, or more precisely the product $Z_hg_0^2$, depends on the UV cutoff $\Lambda$. {\color{black} The radii $\E{r_\mathrm{c}^2}$ and $\E{r_\mathrm{m}^2}$ are proper quantum mechanical observables and thus must be independent of the cutoff up to higher order corrections in the EFT. Since $\E{r_\mathrm{c}^2},\;\E{r_\mathrm{m}^2}\propto Z_hg_0^2$, we must require $Z_hg_0^2$ to depend only on inverse powers of the UV cutoff $\Lambda$, i.e.,\footnote{Note that this equation does not qualify itself as a renormalization condition, since there is an ambiguity due to the unknown constant.}
\begin{equation}\label{eq:Zg02_well_defined}
	(Z_hg_0^2)(\Lambda)\stackrel{!}{=}\const(a,B,A)+\mathcal{O}\left(\Lambda^{-1}\right)
\end{equation}
in order to separately predict the radii.  In the following, we will use the same quantities as renormalization input as in \cite{HongoSon22}, but 
we consider two different strategies  (i) and (ii) to implement the renormalization using one renormalization condition. These efforts are described below. In the end, they will not be successful, leading us to conclude that a second renormalization condition is required. This will be discussed in Sec.~\ref{sec:full_reno}.}

The renormalization input is provided by the three-body binding energy $B$, that already entered into the theory through the form factors \cite{HongoSon22}. For renormalization purposes it can be introduced by demanding that the dressed trimer propagator has a pole at the corresponding energy. Therefore, we have one renormalization condition
up to this point. More precisely, we require that the dressed trimer propagator at rest has a pole at $p_0=-B<0$, i.e., $G_h^{-1}(p_0=-B,\boldsymbol{p}=\boldsymbol{0})=0$. Evaluating this condition, one gets
\begin{equation}\label{eq:halo_reno_condition_pole_final}
    (B_0-B)+4\pi g_0^2I_\Lambda(B)=0\,.
\end{equation}
This renormalization condition can in principle be satisfied, e.g., by letting $B_0$ and $g_0$ run with the UV cutoff\footnote{
We have also carried out the subsequent analysis using dimensional regularization instead of a sharp momentum cutoff. This leads to the same results.}. We have two parameters $g_0$ and $B_0$ but only one condition. Therefore, the system is under-determined. To resolve this, we assume one of the parameter to be independent of $\Lambda$. 

{\color{black} Our first ansatz (i)} to use the renormalization condition assumes that $g_0$ does not depend on $\Lambda$. This directly implies that  $B_0(\Lambda)$ is $\Lambda$-dependent. Otherwise the condition~\eqref{eq:halo_reno_condition_pole_final} cannot be satisfied. But since $B_0$ neither appears explicitly in the radii nor implicitly via $Z_h$ in Eq.~\eqref{eq:field_strength_reno}, we do not need the explicit running of $B_0(\Lambda)$ here. Since we assume $g_0$ to be constant, we just need to know how $I_\Lambda'(B)$ behaves for large $\Lambda$. From the definitions in Eq.~\eqref{eq:integral} one can already see that $I_\Lambda'(B)$ diverges logarithmically as $\Lambda\to\infty$. For a more explicit expression we refer to Appendix~\ref{ap:self-energy}. We conclude
\begin{equation}
    Z_hg_0^2=\frac{1}{g_0^{-2}-4\pi I_\Lambda'(B)}\xrightarrow[\Lambda\to\infty]{}0\,.
\end{equation}
Since $\E{r_\mathrm{c}^2},\;\E{r_\mathrm{m}^2}\propto Z_hg_0^2$, we immediately see that $\E{r_\mathrm{c}^2},\;\E{r_\mathrm{m}^2}\to0$ as $\Lambda\to\infty$. Thus we do not obtain finite radii with {\color{black} ansatz (i)} in the
limit $\Lambda\to\infty$.

{\color{black}In our second ansatz (ii),} we incorporate a running $g_0(\Lambda)$. But in order to arrive there in a unique way, we need the second quantity $B_0$ to be constant. Any choice of a running $B_0(\Lambda)$ cannot be justified without further modifications or requirements. With a constant $B_0$, especially $B_0=0$, we can directly solve for $g_0^2(\Lambda)$ from Eq.~\eqref{eq:halo_reno_condition_pole_final} and express it in terms of the divergent integral:
\begin{equation}
    g_0^2(\Lambda)=\frac{B-B_0}{4\pi I_\Lambda(B)}\,.
\end{equation}
The integral $I_\Lambda(B)$ diverges quadratically in the cutoff $\Lambda$. We then multiply $g^2_0(\Lambda)$ with $Z_h$. In the resulting expression there appears now a difference of the integral $I_\Lambda(B)$ and its derivative each with their respective asymptotic behavior. For large $\Lambda$ the quadratic term dominates. Thus, we get the asymptotic behavior:
\begin{equation}
    Z_hg_0^2={\Big(}\underbrace{\frac{4\pi I_\Lambda(B)}{B-B_0}}_{\sim\Lambda^2}-\underbrace{4\pi I_\Lambda'(B)}_{\sim\ln(\Lambda)}{\Big)}^{-1}\sim\frac{1}{\Lambda^2}\xrightarrow[\Lambda\to\infty]{}0\,.\end{equation}
Thus our second {\color{black}ansatz (ii)} also leads to vanishing radii in the limit $\Lambda \to \infty$. 

We identify the problem as being due to the number of divergences in the EFT. More precisely, we consider the self-energy in this theory from Eq.~\eqref{eq:self-energy} as a function of the Galilei invariant energy $E=-p_0+\tfrac{\boldsymbol{p}^2}{2m_h}$. If we expand the self-energy in some energy $\widetilde{E}$, we get the series
\begin{equation}
    \Sigma_h(E)\propto I_\Lambda(\widetilde{E})+I_\Lambda'(\widetilde{E})(E-\widetilde{E})+\mathcal{O}(\Lambda^{-1})\,.
\end{equation}
This analysis reveals that there are exactly two divergences for $\Lambda\to\infty$. The first one is $I_\Lambda(\widetilde{E})\sim\Lambda^2$, which appears as our renormalization condition where $\widetilde{E}=B$. The second one is $I_\Lambda'(\widetilde{E})\sim\ln(\Lambda)$, which enters through the definition of the field strength renormalization. Higher derivatives of $I_\Lambda(E)$ are convergent and therefore only yield terms of order $\Lambda^{-1}$.

{\color{black} As a consequence, we need two renormalization conditions to absorb all divergences.
In the {\color{black} HS scheme}, there is one renormalization condition which contains two free parameters. One parameter can be used to absorb the first divergence. However, a second renormalization condition is required to predict the radii separately (as well as any other observable).}

\section{Full Renormalization and Landau Pole}\label{sec:full_reno}
Our aim is to arrive at definite predictions for the radii in our EFT. For this purpose one can exploit the fact that all radii
are proportional to $Z_hg_0^2$,  which is exactly the vanishing term from the last section. Thus, it is inviting to determine this expression by adding a second renormalization condition that fixes one of the two radii. Thereby, we can  directly satisfy our requirement in Eq.~\eqref{eq:Zg02_well_defined}. 
We get the following requirement either for the charge or matter radius:
\begin{equation}\label{eq:second_reno_radius}
    \E{r_\mathrm{c/m}^2}=K_\mathrm{c/m}Z_hg_0^2\overset{!}{=}\E{r_\mathrm{c/m}^2}_\mathrm{exp}\,.
\end{equation}
Some value
of either the charge or matter radius is needed as renormalization input\footnote{We can also use the neutron radius or the $E1$ strength in this condition, since they all have the same dependence on $g_0$ and $B_0$.}, which, e.g., can be taken from experimental results. From this equality we can immediately write down the prediction for the respective other radius depending on one radius as input:
\begin{equation}
    \E{r_\mathrm{c}^2}=\frac{K_\mathrm{c}}{K_\mathrm{m}}\E{r_\mathrm{m}^2}_\mathrm{exp}\quad\text{or}\quad\E{r_\mathrm{m}^2}=\frac{K_\mathrm{m}}{K_\mathrm{c}}\E{r_\mathrm{c}^2}_\mathrm{exp}\,.
\end{equation}
Here, the prefactor $K_\mathrm{m}/K_\mathrm{c}$ is equal to the universal ratio of matter and charge radius, Eq.~\eqref{eq:ratio_univ}, predicted by HS. So far, these seem to be unproblematic equations. But if one investigates the adjusted coupling $g_0(\Lambda)$, the full renormalization makes one specialty of this EFT explicit, namely the existence of a  Landau pole. Inserting the expression of $Z_h$ in terms of $g_0$ from Eq.~\eqref{eq:field_strength_reno} into our condition Eq.~\eqref{eq:second_reno_radius}, we arrive at a requirement for the coupling in order to maintain the renormalization condition:
\begin{equation}\label{eq:g0-2}
    \frac{1}{g_0^2(\Lambda)}=\frac{K_\mathrm{c/m}}{\E{r_\mathrm{c/m}^2}_\mathrm{exp}}+4\pi I_\Lambda'(B)\,.
\end{equation}
We first note that while the first term in the expression for $g_0^{-2}(\Lambda)$ is a positive number, the integral $I_\Lambda'(B)$ is always negative and monotonically decreasing with $\Lambda$. For sufficiently large $\Lambda$, the term $I_\Lambda'(B)$ becomes arbitrarily negative, behaving like $I_\Lambda'(B)\sim-\ln(\Lambda)$. Therefore,  
there is an upper limit for the UV cutoff $\Lambda$ above which $g_0^{-2}(\Lambda)$ becomes negative and hence the coupling itself needs to be imaginary to maintain the renormalization condition. This, in turn,  would lead to a non-hermitian Hamiltonian. 
The maximum value of the cutoff $\Lambda_\mathrm{max}$ that preserves hermiticity determines a Landau pole of our EFT. Its numerical value depends on the system parameters. From Eq.~\eqref{eq:g0-2} we can directly see that $\Lambda_\mathrm{max}$ becomes larger when the input radius becomes smaller. With a set of values for all parameters one can in principle determine $\Lambda_\mathrm{max}$ by finding the root of Eq.~\eqref{eq:g0-2}. We will present and discuss values for the Landau pole in different cases in Sec.~\ref{sec:comparison}. Also, we will calculate a three-body scattering amplitude in order to check for the proper renormalization procedure of a different observable in Sec.~\ref{sec:scattering}.

\section{Range of Applicability}\label{sec:comparison}

In this section, we access the applicability of this specific version of the {\color{black} HS scheme}
in two different ways.
First, we compare the matter radii obtained for different two-neutron halos with the ones from standard Halo EFT calculations including the \(nc\) interaction.
Second, we calculate the position of the Landau pole for different nuclei and compare it to the high-momentum scale of the nuclei.
Based on this, we check if the pole restricts the region of validity in practice.

\subsection{Comparison with Halo EFT Three-Body Calculations}

In order to get an impression of the accurateness of the radii calculated in this approach we compare with radii obtained from full Halo EFT calculations.
A central point in standard Halo EFT is describing the three-body dynamics by solving the Faddeev equations for the Faddeev amplitudes \(\ket{F_i}\). In the non-relativistic effective field theory this system of equations is equivalent to the Schrödinger equation.
It reads
\begin{equation}\label{eq:fd}
    \ket{F_i} = \sum_{j \neq i} G_0 t_j \ket{F_j} \,,
\end{equation}
whereby \(G_0\) is the free Green's function of the three-body system and \(t_j\) are the two-body t-matrices corresponding to the potentials \(V_j\) appearing in the Schrödinger equation.
In the EFT treatment the t-matrices can be directly parameterized in terms of the effective-range expansion parameters.

In practice, one has to work with a basis representation of Eq.~\eqref{eq:fd}.
A typical choice is a Jacobi basis of two momentum vectors \(p\) and \(q\) with respect to some spectator \(i\) (\(n\), \(n'\), or \(c\)).
The relations for the momenta read
\begin{align}
    \vec{p}_i &\coloneqq \mu_{jk} \left( \frac{\vec{k}_j}{m_j} - \frac{\vec{k}_k}{m_k} \right)  \,, \\
    \vec{q}_i &\coloneqq \mu_{i(jk)} \left( \frac{\vec{k}_i}{m_i} - \frac{\vec{k}_j + \vec{k}_k}{M_{jk}} \right)  \,,
\end{align}
whereby \(\mu_{jk} = m_j m_k /\K{m_j + m_k}\), \(M_{jk} = m_j + m_k\), and \(\mu_{i(jk)} = m_i M_{jk} / \K{m_i + M_{jk}}\) hold.
Once the Faddeev equations are represented by functions after applying a basis, one obtains a set of
coupled integral equations.
These can be solved numerically. Discretizing the integrals turns it in an eigensystem.
In addition to the two-body forces, a three-body force is used to renormalize the three-body system
to the correct \(B\).
From the solutions for the Faddeev amplitudes also the full wave function can be obtained.
The underlying relation therefor is given by 
\begin{equation}
    \ket{\Psi} = \sum_i G_0 t_i \ket{F_i} \,.
\end{equation}
As the single summands are typically expressed with respect to the Jacobi basis in which the matrix elements of \(t_i\) and the representation of \(\ket{F_i}\) is known, one has to carry out 
a recoupling to the Jacobi basis (corresponding to a different spectator) in which one chooses to represent the overall state in for some of the terms.
For a more detailed discussion of this variation of the Faddeev formalism, we refer to Refs.~\cite{Gobel:2019jba,Gobel:2022pvz} and for a more generic approach to the recoupling to Refs.~\cite{Kirchner:2024axc,Gobel:2024ovk}.

We obtain the mean square radii \(\expval{r_c^2}\) and \(\expval{r_n^2}\) from the derivative of the corresponding form factors.
The form factors \(\mathcal{F}_i{(\vec{k})}\) (\(i \in \{n, c\}\)) are given by
\begin{align}\label{eq:F_i_def}
    \mathcal{F}_i{(\vec{k})} &\coloneqq \expval{P_{\vec{k}}^{(i)}} = \bra{\Psi} \K{ \int \dd^3{\vec{p}} \int \dd^3{\vec{q}} \iketbrae{\vec{p},\vec{q}}{i}{\vec{p},\vec{q}+\vec{k}} \otimes \mathbb{1}^{(\mathrm{spin})} } \ket{\Psi} \\
    &= \int \dd^3{\vec{p}} \int \dd^3{\vec{q}} \sum_{\xi} \ibraket{}{\Psi}{\vec{p},\vec{q};\xi}{i}
      \ibraket{i}{\vec{p},\vec{q}+\vec{k};\xi}{\Psi}{} \,, \label{eq:F_i_2}
\end{align}
whereby \(P_{\vec{k}}^{(i)}\) is the operator that shifts the momentum \(\vec{q}\) by \(\vec{k}\).
In the second row, the multiindex containing the spin quantum numbers is given by \(\xi\).
On this basis, one can define an angular averaged variant given by
\begin{equation}\label{eq:F_i_aavg}
    \widetilde{\mathcal{F}}_i{(k^2)} \coloneqq \frac{1}{4\pi} \int \dd{\Omega_{\vec{k}}} \mathcal{F}_i{(\vec{k})} \,.
\end{equation}
The radii \(\expval{y_i^2}\) are then related to the derivative of the form factor in \(k^2\) at zero:
\begin{equation}\label{eq:y_i_deriv}
    \expval{y_i^2} = \K{-6} \partial_{k^2} \widetilde{\mathcal{F}}_i{(k^2)} \big|_{k^2=0} \,.
\end{equation}
Note that \(y_i\) is the coordinate-space correspondent to the momentum \(q_i\).
The conversion factors to the \(r_i\) coordinates are given by \(r_c = 2/\K{A+2} y_c\) and \(r_n = \K{A+1}/\K{A+2} y_n\).
If the radii are obtained in this way the contributions from different partial-wave components of the halo bound state \(\ket{\Psi}\) are included.
In practice, the expression for the angular-averaged form factor is evaluated in a partial-wave basis with a truncation in the partial waves.
Therefore, the contributions of the most important partial waves are included.

We perform the comparison for the two-neutron halo nuclei \(^{11}\)Li, \(^{14}\)Be, \(^{17}\)B, \(^{19}\)B, \(^{22}\)C as well as for \(^6\)He.
The latter differs from the other halos, because  it has a p-wave interaction at leading order.
The leading-order component of the \(n\alpha\) interaction is in the \(^2P_{3/2}\) partial wave.
While in the case of \(^{14}\)Be and \(^{22}\)C the halo nuclei as well as their core have spin zero, this is not true for \(^{11}\)Li, \(^{17}\)B, and \(^{19}\)B.
However, for these nuclei the overall spin and core spin \(s_c\) are still the same.
If we assume that the \(nc\) interaction is the same in the \(j=s_c+1/2\) and in the \(j=s_c-1/2\) channel, we can neglect the core spin as well as the overall spin.
For details see the discussion in Ref.~\cite{Gobel:2022pvz}.
For the calculation of the s-wave halo nuclei, we use the parameters listed in Tab.~\ref{tab:sw_halos_param}.

\begin{table}[H]
  \centering
  \caption{Core mass numbers, two-neutron separation energies \(B\), and \(nc\) virtual-state energies \(E_{nc}^*\) of the considered s-wave two-neutron halo nuclei.
    The values for \(B\) and for \(E_{nc}^*\) are
    taken from Ref.~\cite{Wang:2021xhn}.
    The only exception is  \(^{22}\)C, for which we use the \(B\) value
    given in Ref.~\cite{Hammer:2017tjm}. The virtual-state energy \(E_{nn}^*\)  corresponds to a neutron-neutron scattering length of -18.7~fm \cite{GonzalezTrotter:1999zz} and is the same in all cases.
  }\label{tab:sw_halos_param}
  \begin{tabular}{lcccc}
    \toprule
      nucleus & \(A\) & \(B/\)keV & \(E_{nc}^*/\)keV & \(E_{nn}^*/\)keV \\
    \midrule
      \(^{11}\)Li & 9  &  369 &  26 & \\
      \(^{14}\)Be & 12 & 1266 & 510 & \\
      \(^{17}\)B  & 15 & 1384 &  83 & 118.5 \\
      \(^{19}\)B  & 17 &   90 &   5 & \\
      \(^{22}\)C  & 20 &  100 &  68 & \\
    \bottomrule
  \end{tabular}
\end{table}

For \(^6\)He, we use a \(B\) of 975~keV \cite{Brodeur:2011sam}.
The \(nc\) t-matrix is parameterized via its reduced t-matrix element given by
\(\tau_{nc}{(E)} = \K{4\pi^2 \mu_{nc}}^{-1} \K{\gamma_1 \K{k^2 - k_R^2}}^{-1}\) with
\(\gamma_1 = -r_1/2\) and \(k_R = \sqrt{2/(a_1 r_1)}\) \cite{Bedaque:2003wa,Ji:2014wta}.
We employ the parameters \(r_1=-174.0227\)~MeV and \(k_R=37.4533\)~MeV (based on the \(a_1\) and \(r_1\) values from Ref.~\cite{Arndt:1973ssf}).
While in the calculations of the s-wave halo nuclei we used the standard implementation of Halo EFT with t-matrices corresponding to zero-range interactions, for \(^6\)He we used for the \(n\alpha\) interaction a t-matrix corresponding to a finite-range interaction.
This is to avoid difficulties stemming from energy-dependent interactions, as the p-wave zero-range interaction would be energy-dependent.
For more details on this issue see Ref.~\cite{Gobel:2019jba}, while the finite-range version of Halo EFT will be discussed in Ref.~\cite{fr_heft}.

\begin{figure}[H]
    \centering
    \includegraphics[width=0.9\linewidth]{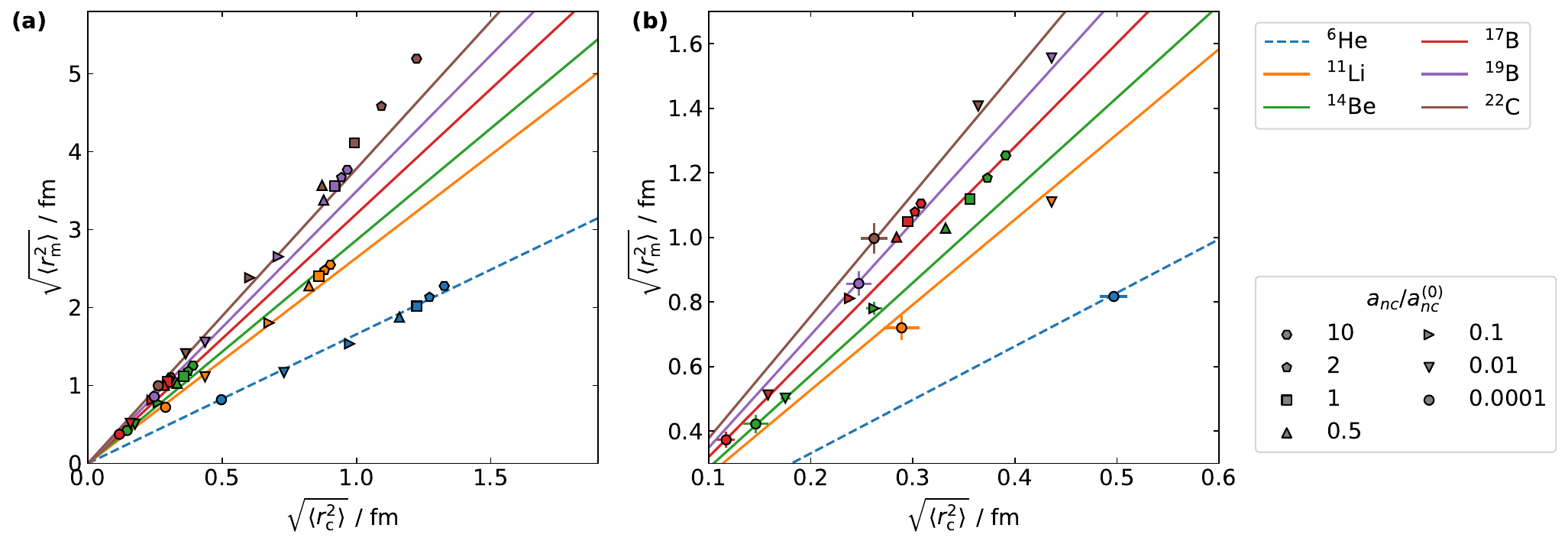}
    \caption{The lines show the HS results for the relation between RMS matter and charge radius for the different halo nuclei.
    Except of \(^6\)He with the LO \(nc\) interaction in a p-wave, all other halos have the \(nc\) interaction in the s-wave.
    This specialty is indicated by a dashed line in the case of \(^6\)He.
    The symbols depict results from standard Halo EFT. The color encodes the nucleus using the same color scheme as the lines, while the shape encodes the ratio of the actually used \(nc\) scattering length \(a_{nc}\) and the physical \(nc\) scattering length \(a_{nc}^{(0)}\).
    The right panel \textcolor{black}{(b)} shows a zoomed-in version of the left panel \textcolor{black}{(a)}.
    Error bars depicting estimated uncertainties related to numerics and the extraction of the radii are also included. In most cases, these are too small to be visible.}
    \label{fig:radii_calculation}
\end{figure}

In order to compare the results from standard Halo EFT with the prediction of {\color{black} the HS scheme},
we have calculated for these nuclei the radii for different rescalings of the \(nc\) scattering length \(a_{nc}\).
Thereby, we can in the theory go artificially closer to the limit of a negligible \(nc\) interaction.
We specify the rescaling as the ratio \(a_{nc} / a_{nc}^{(0)}\), whereby \(a_{nc}^{(0)}\) is the physical value of the \(nc\) scattering length in the corresponding system.
In Fig.~\ref{fig:radii_calculation}, we show the matter radius as a function of the core radius.
The solid lines show the predictions {\color{black} of the HS scheme} for the different nuclei.
The data points depict the results from standard Halo EFT, whereas the color encodes
the nucleus and the shape encodes the rescaling of the \(nc\) scattering length.
The left and right panel just differ in the plotting region.

We observe that the results converge to the prediction of {\color{black} the HS scheme} if \(a_{nc}\) is artificially lowered.
Moreover, the plots show that for artificially increased \(nc\) scattering lengths the deviations from HS become larger.
At the physical \(nc\) scattering length, \(^6\)He is the nucleus with the smallest deviation of the matter radius from the HS prediction. The relative deviation amounts to only 0.6~\%.
In the following we discuss the s-wave halos in more detail.
Among them, \(^{11}\)Li displays the smallest deviation of its matter radius from the HS prediction. The relative deviation is 5.9~\%.
The nuclei \(^{14}\)Be and \(^{22}\)C display deviations of about 9.7~\%.
The largest deviations are displayed by \(^{19}\)B and \(^{17}\)B which show deviations of 11.0~\% and 11.1~\%.
Given that the requirement of applicability of {\color{black}the HS scheme} is not strictly fulfilled, these are quite good results.
(For the Borromean halo nuclei considered here, the theory requires \(E_{nc}^{*}\) to be large compared to the binding energy of three-body system and the \(nn\) virtual state energy.)
However, for the physical \(nc\) scattering length this is not the case.
The ratio \(E_{nc}^{*} /B\) is for the considered Boron halos about 6~\%, for \(^{11}\)Li about 7~\%, and for \(^{14}\)Be around 40~\%, while for \(^{22}\)C it amounts to 68~\%.
Correspondingly, for a rescaling of the \(nc\) scattering length by 0.1 the ratios are already a factor 100 larger. 
Even for the two Boron halos \(E_{nc}^{*}\) is then about six times larger than the binding energy and for the other halos the separation is even larger.
We notice that for the rescaling with 0.1 the data points are already quite close to the HS universal curves.
The relative deviations are now below 8~\% and if we exclude the Boron isotopes they are even below 5~\%.
For even smaller \(nc\) scattering length the agreement is even better.
For a rescaling of 0.0001 the standard Halo EFT results lie on top of the HS curves.
The halo \(^{11}\)Li is a little bit off, but within the uncertainty estimate it is consistent with the HS curve.
For a visualization of the relative deviations in percent see Appendix~\ref{ap:radii_rel_devs} and the figure therein.
The figure contains also error bars specifying uncertainty estimates based on extraction (from the form factor) and numerical (number of grid points, cutoff) uncertainties.
In conclusion, the numerical calculations show that the HS results can be obtained as a limit of standard Halo EFT.
This is in line with the analytical findings of Naidon in Ref.~\cite{naidon}.
However, we note that Naidon used a different limit to approach the {\color{black}HS scheme} and did not include an explicit three-body force. In Ref.~\cite{naidon}, the \(nc\) interaction was varied to reproduce the binding energy of the three-body system.
In contrast, we use a three-body force to fix the binding energy of the three-body system while the \(nc\) interaction strength is fixed by the properties of the physical \(nc\) subsystem.

Fig.~\ref{fig:radii_calculation} also shows nicely that decreasing \(a_{nc}\), the matter radius does not only get closer to the HS curve but also the charge radius changes.
Thereby, the points move more to the origin, to smaller matter and charge radii.
Except for \(^6\)He and \(^{11}\)Li we observe that all points converge from above against the universal curve.
Taking our estimates of numerical and extraction uncertainties (see, e.g., Fig.~\ref{fig:radii_rel_devs}) at face value, the different behavior of \(^6\)He and \(^{11}\)Li is not a numerical artifact.
We note that the estimated uncertainties tend to be larger for the rescaling factors of 0.01 and 0.0001 for all nuclei.
Due to the smaller spatial extensions, these momentum-space calculations are more challenging for these factors.
For a discussion of the uncertainty estimation and for plots of the absolute values with uncertainty estimates see Appendix~\ref{ap:add_info_radii}.
A detailed discussion of the relative deviations and the convergence behavior can be found in Appendix~\ref{ap:radii_rel_devs}.

It is additionally interesting to note that the HS approach was also compared to standard Halo EFT at the example of the \(E1\) strength distribution which parametrizes the nuclear structure contribution to Coulomb dissociation cross section in Ref.~\cite{Gobel:2022pvz}.
When comparing the distribution with \(nn\) final-state interactions included significant deviations were found amounting to approximately a factor of 3 at the low-energy peak.
However, it is worth mentioning that the coupling constant \(g\) was not determined exactly for this process but estimated as described in Ref.~\cite{HongoSon22} in the context of computing \(r_c\) of \(^{11}\)Li.
It is also important to note that \(r_c^2\) and the integral over the \(E1\) strength distribution are connected via the non-energy weighted sum rule.
Given that we obtain an \(r_c\) of 0.87~fm or 0.86~fm here, if we include only the s-wave, and that HS reported a radius of 0.86~fm using the approximate coupling $g$ this discrepancy might seem a bit counterintuitive.
However, two aspects have to be taken into account here. First, the sum rule constraints the overall area under the curve of the \(E1\) distribution but not its shape. Second, we use a value of 369~keV
for the two-neutron separation energy, while HS used 247~keV.
For the comparison of the \(E1\) distributions  the HS \(E1\) distribution was evaluated at $B=\SI{369}{keV}$ in Ref.~\cite{Gobel:2022pvz}, using their estimation method for the coupling $g$.
With this input value for \(B\), {\color{black} the HS Halo EFT scheme} produces a \(r_c\) of about 0.75~fm.
Then the areas under the \(E1\) curves are approximately different by a factor of \((0.75/0.86)^2 \approx 0.76\).
In summary, the charge radius from Halo EFT and {\color{black} HS scheme} is quite different when computed with consistent parameters. This is in line with the findings of Ref.~\cite{Gobel:2022pvz}. However, if our renormalization procedure is applied, then the discrepancy between the radii is significantly smaller.

\subsection{Landau Pole}
As the UV momentum cutoff $\Lambda$ approaches the maximum cutoff $\Lambda_\mathrm{max}$, the squared coupling $g_0^2(\Lambda)$ becomes infinitely large, since in Eq.~\eqref{eq:g0-2} $g_0^{-2}(\Lambda)$ becomes zero. Thus the dimer-core coupling $g_0(\Lambda)$ runs to a Landau pole in the UV. To state an explicit value for $\Lambda_\mathrm{max}$ we need to specify our system, i.e., we need to fix the mass $A$ of the core and the two-neutron separation energy $B$. Further we need some radius as input for our renormalization. For illustrative purposes, we consider \ce{^22C} since 
the matter radius of this halo nucleus has been measured \cite{tanaka_c22,togano:2016}. We also consider other 2n halo nuclei and  use radii from our Halo EFT calculation to calculate values of the Landau pole in each case respectively. {\color{black} We compare the so-found values with estimates of the low-momentum scales at which we assume the relative momenta to be located as well as with the estimated breakdown scales.} First there is the low-momentum scale in the two-neutron subsystem, which is given by $M_\mathrm{low}^\mathrm{nn}=1/|a_\mathrm{nn}|
\approx\SI{11}{MeV}$ independent of the considered nucleus. The value of the nn scattering length $a_\mathrm{nn}=-\SI{18.7}{fm}$ is taken from Ref.~\cite{GonzalezTrotter:1999zz} and is used for the subsequent computations. The second scale is that of the dimer-core system, which is given by $M_\mathrm{low}^\mathrm{nnc}=\sqrt{2\mu_\mathrm{dc}B}$, where $\mu_\mathrm{dc}$ is the reduced mass of the dimer-core system. On the other hand we can identify three breakdown scales. In the two-neutron system this is set by the nn effective range to $M_\mathrm{high}^\mathrm{nn}=1/|r_\mathrm{nn}|
\approx\SI{72}{MeV}$, where we took the value $r_\mathrm{nn}\approx r_\mathrm{np}=\SI{2.73(3)}{fm}$ from Ref.~\cite{preston1975structure} assuming isospin symmetry. The two other breakdown scales are set by core excitation to $M_\mathrm{high}^\mathrm{c}=\sqrt{2\mu_\mathrm{dc}E_\mathrm{c}^*}$ and for the neutron-core scattering to $M_\mathrm{high}^\mathrm{nc}=|a_\mathrm{nc}^{-1}|$, where if not given otherwise, we estimate $|a_\mathrm{nc}^{-1}|\approx\sqrt{2\mu_\mathrm{nc}E_\mathrm{nc}^*}$ using the neutron-core virtual-state energy $E_\mathrm{nc}^*$ and reduced neutron-core mass $\mu_\mathrm{nc}$. {\color{black} The effects of the neutron-core interaction are nominally suppressed by $M_\mathrm{low}/M_\mathrm{high}^\mathrm{nc}$. For some of the examples discussed below, this factor is not small, which indicates that the HS scheme is not applicable. We still consider these nuclei for illustrative purposes.} For the first excitation energy $E^*_\mathrm{c}$ of the core we take values from Refs.~\cite{levels_c20, ex_state_he4, ex_states_li9, ex_states_be12, ex_states_b15, ex_states_b17}.

\begin{table}[h]
    \begin{center}
        \caption{Landau pole for $\ce{^{22}C}$ for selected values of $B$ and $\sqrt{\expval{r_\mathrm{m}^2}}$.
        } 
        \label{tab:22c}
        \begin{tabular}{c|cccccccc}
            $\sqrt{\expval{r_\mathrm{m}^2}}$/\si{fm} & $3.36$ & $6.3$ & $3.36$ & $6.3$ & $3.36$ & $6.3$ & $3.36$ & $6.3$ \\
            \midrule
            $B/\si{keV}$ & $1$ & $1$ & $10$ & $10$ & $50$ & $50$ & $100$ & $100$ \\
            \midrule
            $\Lambda_\mathrm{max}/\si{MeV}$ & $142$ & $33$ & $61$ & $22$ & $39$ & $19$ & $36$ & $19$ \\
        \end{tabular}
    \end{center}
\end{table}
For \ce{^22C} we fix the matter radius. There is a value of $\SI{5.4(9)}{fm}$ extracted via a calculation from a reaction cross section measurement~\cite{tanaka_c22} as well as a newer value of $\SI{3.44(8)}{fm}$ from another measurement~\cite{togano:2016}. Since the coupling $g_0^{-2}(\Lambda)$ depends monotonically on the radius input, we consider both the larger and smaller value including their uncertainty. Ref.~\cite{acharya_c22} relates the matter radius and the neutron-core virtual-state energy to an upper bound of $B$, which in our case implicates $B<\SI{100}{keV}$. We take this upper bound as our value for $B$. By calculating the root of Eq.~\eqref{eq:g0-2} numerically, we get the results in Tab.~\ref{tab:22c}. For comparison, the EFT scales for \ce{^22C} are listed in Tab.~\ref{tab:22c_scales}.

\begin{figure}[H]
    \centering
    \includegraphics[width=0.7\linewidth]{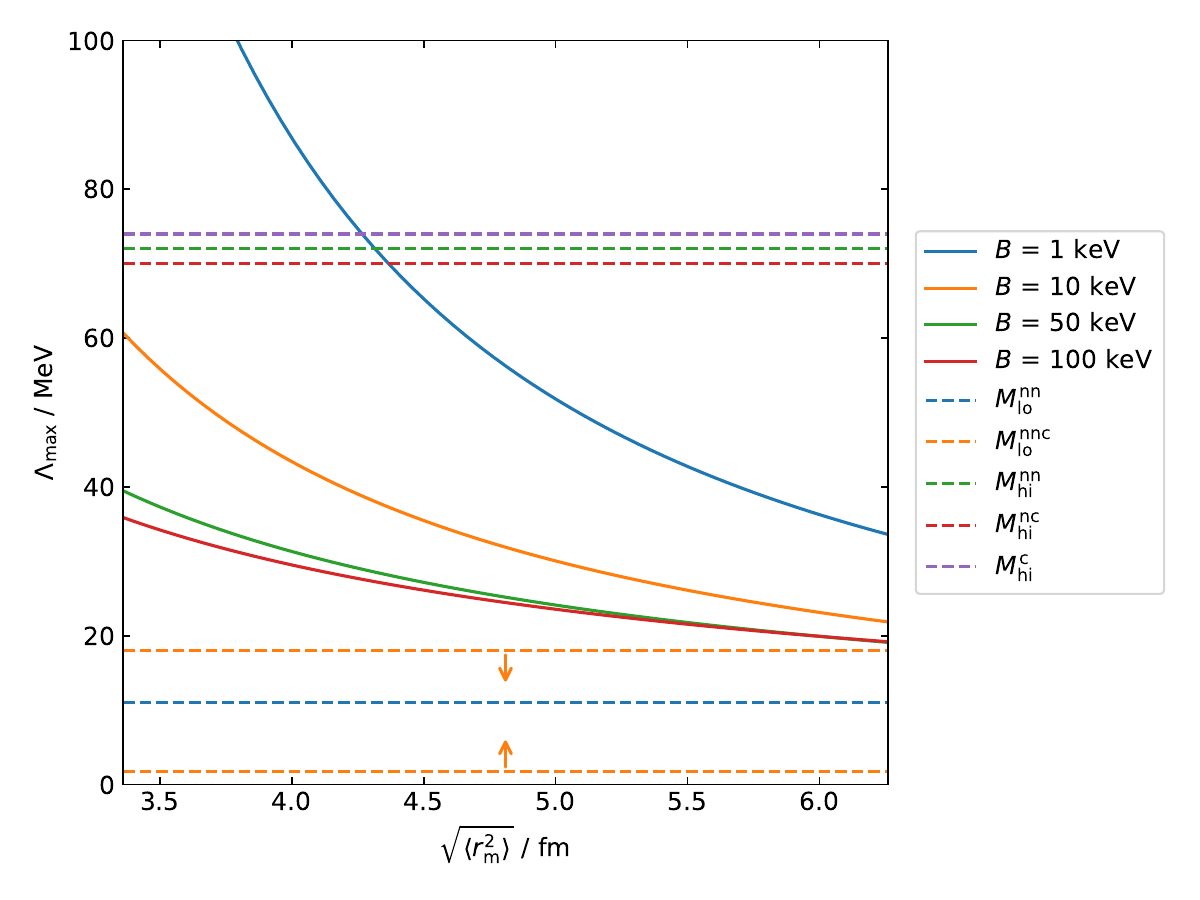}
    \caption{Landau pole values for \ce{^22C} as a function of $\sqrt{\E{r^2_\mathrm{m}}}$ for different values of $B$. The EFT scales for this case are added as horizontal lines. Note that $M^\mathrm{nnc}_\mathrm{lo}$ depends on $B$. We indicate this with two horizontal lines at $B=\SI{100}{keV}$ and $B=\SI{1}{keV}$ as an upper and lower bound, respectively.}
    \label{fig:landau_plot}
\end{figure}

\begin{table}[h]
    \begin{center}
    \caption{EFT scales for $\ce{^{22}C}$ in MeV in {\color{black} the HS scheme}. We have $\mu_\mathrm{dc}=1.8m_\mathrm{n}$, $E_\mathrm{c}^*=E^*(\ce{^20C})=\SI{1.6}{MeV}$ and $|a_\mathrm{nc}|<\SI{2.8}{fm}$~\cite{mosby_c22}.} 
    \label{tab:22c_scales}
    \begin{tabular}{cc|ccc}
        $M_\mathrm{lo}^{\mathrm{nn}}$ & $M_\mathrm{lo}^\mathrm{nnc}$ & $M_\mathrm{hi}^\mathrm{nn}$ & $M_\mathrm{hi}^\mathrm{nc}$ & $M_\mathrm{hi}^\mathrm{c}$ \\
        \midrule
        $10$ & $\lesssim 18$ & $72$ & $\gtrsim 70$ & $74$ \\
    \end{tabular}
    \end{center}
\end{table}
We find a Landau pole ranging from $\SI{19}{MeV}$ to $\SI{142}{MeV}$, depending on the input values for $\sqrt{\E{r^2_\mathrm{m}}}$ and $B$. {\color{black}Additionally in Fig.~\ref{fig:landau_plot}, we plot the Landau pole for selected values of $B$ as a function of the matter radius together with the EFT scales to see the behavior of the Landau pole in more detail. For values $B=\SI{50}{keV}$ and $\SI{100}{keV}$, the Landau pole is quite low being located around $\SI{30}{MeV}$. 
This is much lower than the high-momentum scales of the EFT around 70 MeV and close to the low-momentum scales located around $\SI{10}{}$ to $\SI{20}{MeV}$. Thus the range of applicability of the EFT is significantly restricted. For a low value of $B=\SI{1}{keV}$ the Landau pole shows more variation exceeding the high-momentum scales for small radius.} The huge range of these values is mostly due to the large uncertainty of the binding energy and the matter radius. A small sensitivity is found from the uncertainty of the neutron-neutron scattering length. Since the Landau pole ranges from near the low-momentum scales up to above the high-momentum scales, we cannot make any definite statements on this issue. But in certain scenarios this EFT might suffer from the relatively small value of its Landau pole, thus restricting its naive range of applicability.

For the other 2n halo nuclei, we use the input values of $B$ and $E_\mathrm{nc}^*$ from our standard Halo EFT calculation to compute the EFT scales. For the radii we take the values from the calculation itself. We determine the Landau poles in the same manner but for both the matter and charge radii per nucleus. All values of Landau poles and EFT scales for these nuclei are collected in Tab.~\ref{tab:2n_halos_landau}.
\begin{table}[H]
  \centering
  \caption{Landau poles of 2n halo nuclei for each radius together with respective EFT scales in {\color{black}the HS scheme}. Note that for the p-wave halo \ce{^6He} we take the values from Ref.~\cite{Bertulani:2002sz}. All quantities are given in \si{MeV}.}
  \label{tab:2n_halos_landau}
  \begin{tabular}{lccccccc}
    \toprule
      nucleus & \(\Lambda_\mathrm{max}(\sqrt{\E{r_\mathrm{m}^2}})\) & \(\Lambda_\mathrm{max}(\sqrt{\E{r_\mathrm{c}^2}})\) & \(M^\mathrm{nn}_\mathrm{lo}\) & \(M^\mathrm{nnc}_\mathrm{lo}\) & \(M^\mathrm{nn}_\mathrm{hi}\) & \(M^\mathrm{nc}_\mathrm{hi}\) & \(M^\mathrm{c}_\mathrm{hi}\) \\
    \midrule
      \(^{6}\)He  & 100 & 99 & & 49 & & 260 & 225 \\
      \(^{11}\)Li & 68 & 79 & & 34 & & 7 & 92 \\
      \(^{14}\)Be & 135 & 156 & 10 & 64 & 72 & 30 & 82 \\
      \(^{17}\)B  & 129 & 150 & & 68 & & 12 & 66\\
      \(^{19}\)B  & 37 & 42 & & 17 & & 3 & 61 \\
    \bottomrule
  \end{tabular}
\end{table}
Overall, the values show a similar range as in the case of \ce{^22C}, however,  they are larger than the low-momentum scales in each case.
On the one hand, it becomes clear that the 2n halo nuclei \(^{11}\)Li, \(^{14}\)Be, \(^{17}\)B, and  \(^{19}\)B are not good candidates for {\color{black}the HS scheme} as the scale \(M^\mathrm{nc}_\mathrm{hi}\) is quite low, indicating a strong neutron-core interaction. On the other hand, 
\(^6\)He might lie in the range of applicability
since the dominant $nc$ interaction is in the p-wave.
Nevertheless, comparing the scales for all nuclei in Tab.~\ref{tab:2n_halos_landau} to the values of the corresponding Landau poles, allows insights on the applicability on a technical level.  While the Landau pole places a restriction on the range of applicability in the case of \(^6\)He, it is clearly irrelevant for \(^{11}\)Li, \(^{14}\)Be, \(^{17}\)B, and  \(^{19}\)B. But for the latter nuclei, the {\color{black}HS scheme} is not applicable in the first place. We conclude that for the relevant examples 
\(^{22}\)C and \(^6\)He, the Landau indeed leads to a reduction of the naive range of applicability.

\section{Trimer Propagator and Scattering}\label{sec:scattering}
\begin{figure}[h]
    \centering
    \begin{tikzpicture}
        \draw [double distance=1.5] (-1,0) -- (1,0);
        \draw [fill=gray] (0,0) circle (.35);
        
        \draw (-2,-1) -- (-1,0) node[pos=.5, scale=.8, rotate=45]{\midarr};
        \node at (-2.8,-1) {c: $(E_\mathrm{c},\boldsymbol{q})$};
        \draw [dashed] (-1.5,0.5) -- (-1,0);
        \draw (-2,1) -- (-1.5,0.5) node[pos=.5, scale=.8, rotate=-45]{\midarr};
        \node at (-3.3,1) {n$\uparrow$: $(E_\uparrow,\boldsymbol{p}-\tfrac{\boldsymbol{q}}{2})$};
        \draw (-2,0) -- (-1.5,0.5) node[pos=.5, scale=.8, rotate=45]{\midarr};
        \node at (-3.45,0) {n$\downarrow$: $(E_\downarrow,-\boldsymbol{p}-\tfrac{\boldsymbol{q}}{2})$};
        
        \draw (2,-1) -- (1,0) node[pos=.5, scale=.8, rotate=-45]{\midarr};
        \node at (2.9,-1) {c: $(E_\mathrm{c}',\boldsymbol{q}')$};
        \draw [dashed] (1.5,0.5) -- (1,0);
        \draw (2,1) -- (1.5,0.5) node[pos=.5, scale=.8, rotate=45]{\midarr};
        \node at (3.45,1) {n$\uparrow$: $(E_\uparrow',\boldsymbol{p}'-\tfrac{\boldsymbol{q}'}{2})$};
        \draw (2,0) -- (1.5,0.5) node[pos=.5, scale=.8, rotate=-45]{\midarr};
        \node at (3.6,0) {n$\downarrow$: $(E_\downarrow',-\boldsymbol{p}'-\tfrac{\boldsymbol{q}'}{2})$};
    \end{tikzpicture}
    \caption{\label{fig:nnc_scattering}Diagram for the scattering amplitude in the center-of-mass frame for neutron-neutron-core scattering. The gray blob denotes the dressed trimer propagator and is evaluated at rest.}
\end{figure}
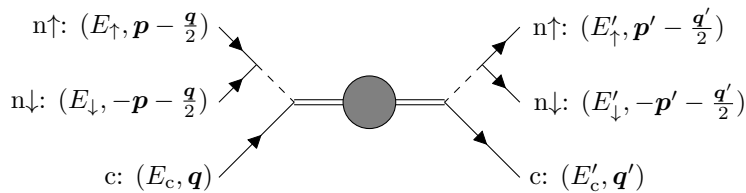

We now consider the dressed halo propagator in more detail as a check of the proper renormalization. Starting from Eq.~\eqref{eq:bare_trimer_prop}, the first step is the elimination of $B_0$ by solving for $B_0=B-4\pi g_0^2 I_\Lambda(B)$ using our renormalization condition. This yields
\begin{equation}\label{eq:G_h_without_B0}
    G_h^{-1}(p)=p_0-\frac{\boldsymbol{p}^2}{2m_h}+B+\i\eps+4\pi g_0^2[I_\Lambda(-p_0+\tfrac{\boldsymbol{p}^2}{2m_h}-\i\eps)-I_\Lambda(B)]\,.
\end{equation}
A cutoff dependence is present via the Integrals $I_\Lambda$. We show using a series expansion of $I_\Lambda(E)$ in Appendix~\ref{ap:self-energy}, that the difference of the integrals depends on $\Lambda$ with $\ln(\Lambda)$ being the leading term. Moreover,  a cutoff dependence can enter through $g_0$ into the propagator. At this point one can already check the consistency of the imaginary part with the effective-range expansion in the hypothetical case of a bound dineutron. This is verified in Appendix~\ref{ap:imag_part_check}. We now proceed to three-body continuum observables and proceed by inserting $g_0^2(\Lambda)$ from Eq.~\eqref{eq:g0-2}.
We are interested in the expression for the scattering amplitudes for a three-to-three neutron-neutron-core scattering for which we can give an analytical expression.
The desired scattering amplitude is obtained by evaluating the diagram depicted in Fig.~\ref{fig:nnc_scattering}.

Since we have no explicit loops anymore, we can just write down the expression for the scattering amplitude\footnote{More accurately, the diagram yields the expression for the t-matrix elements. But these are related to the scattering amplitudes by a simple factor. Note also that we suppressed the dependence on energies and momenta in our notation.}
 using our Feynman rules. Energies and momenta are chosen in the center-of-mass frame. Since the dimer propagators are explicitly known and well-behaved, the only potentially problematic factor in this expression is the dressed trimer propagator with a breakup and subsequent formation $g_0^2\, G_h(E_\mathrm{cm},\boldsymbol{0})$, where $E_\mathrm{cm}:=E_\uparrow+E_\downarrow+E_\mathrm{c}$. We use the expression for $G_h$ from Eq.~\eqref{eq:G_h_without_B0}, where $B_0$ is eliminated, and our second renormalization condition from Eq.~\eqref{eq:g0-2} to eliminate $g_0$ in $G_h$ in favor of $\E{r_\mathrm{c/m}^2}_\mathrm{exp}$. We find
\begin{equation}\label{eq:scattering_finite}
    \left(g_0^2 G_h(E_\mathrm{cm},\boldsymbol{0})\right)^{-1}=\frac{K_\mathrm{c/m}}{\E{r_\mathrm{c/m}^2}_\mathrm{exp}}(E_\mathrm{cm}+B)+4\pi\left[(E_\mathrm{cm}+B)I_\Lambda'(B)+I_\Lambda(-E_\mathrm{cm}-\i\eps)-I_\Lambda(B)\right].
\end{equation}
Now we only have to check whether the term in square brackets in Eq.~\eqref{eq:scattering_finite} is independent of $\Lambda$ up to corrections of order $\Lambda^{-1}$. A consideration of the respective analytical expressions confirms this (see Appendix~\ref{ap:reno_scat_amp} for details). Thus, the renormalization procedure is consistent and we are able to predict the scattering amplitude. 

The full expression of the scattering amplitude evaluates to
\begin{equation}\label{eq:scat_amp_nnc}
    t(\mathrm{nnc}\to\mathrm{nnc})=D(E_\uparrow+E_\downarrow,-\boldsymbol{q})D(E_\uparrow'+E_\downarrow',-\boldsymbol{q}')\,g_0^2\,G_h(E_\mathrm{cm},\boldsymbol{0})\,.
\end{equation}
As a final check, the scattering amplitude can also be used as an alternative second renormalization condition. Strictly speaking, we should start with the cross section at some kinematical point for our scattering process as an observable and find a corresponding expression for the scattering amplitude $t_\mathrm{exp}$, which we use as input for renormalization. Since Eq.~\eqref{eq:scat_amp_nnc} constitutes an expression in terms of bare parameters, we use this equation as our second renormalization condition to fix the coupling $g_0$. Again, inserting the expression for $G_h$ from Eq.~\eqref{eq:G_h_without_B0} with $B_0$ eliminated, we arrive at
\begin{equation}\label{eq:f_B0_eliminated}
t_\mathrm{exp}=d\,\left(\frac{\widetilde{E}+B}{g_0^2}+4\pi\left[I_\Lambda(-\widetilde{E}-\i\eps)-I_\Lambda(B)\right]\right)^{-1}\,,
\end{equation}
where we abbreviated $d:=D(E_\uparrow+E_\downarrow,-\boldsymbol{q})D(E_\uparrow'+E_\downarrow',-\boldsymbol{q}')$. Now we just need to solve for $g_0^{-2}(\Lambda)$, obtaining
\begin{equation}\label{eq:scat_g0-2}
\frac{1}{g_0^2(\Lambda)}=\frac{1}{\widetilde{E}+B}\left(\frac{d}{t_\mathrm{exp}}-4\pi\left[I_\Lambda(-\widetilde{E}-\i\eps)-I_\Lambda(B)\right]\right)\,.
\end{equation}
To see whether the same behavior for large $\Lambda$ occurs as before, we consider the analytical expression of $I_\Lambda(-\widetilde{E}-\i\eps)-I_\Lambda(B)$
(see Appendix~\ref{ap:reno_scat_amp} for details). Indeed, the leading term is $\ln(\Lambda)$ with a positive sign in front. According to Eq.~\eqref{eq:scat_g0-2} this implies the existence of a Landau pole in the same manner as before. Thus, the qualitative findings for this renormalization condition agree with the case of radii.

\section{Conclusion and Outlook}
\label{sec:concl}

In this paper we have investigated the renormalization of the {\color{black}Halo EFT scheme} introduced by HS \cite{HongoSon22} for two-neutron halo nuclei with subleading neutron-core interaction.
In the standard variant, the coupling for the breakup of the trimer into the core and dineutron dimer is undetermined.
Therefore, charge and matter radii or the \(E1\) strength distribution can not independently be predicted as function of the neutron-neutron virtual state energy, the binding energy of the halo, and the mass number of its core.
Instead, the ratio of the two radii and the shape of the distribution is predicted.
We investigated different renormalization approaches to overcome this issue.
Strategies like letting the discussed coupling having no cutoff-dependency or choosing the bare three-body energy parameter to be independent of the cutoff did not turn out to be fruitful.
They yield vanishing radii in the limit of the cutoff \(\Lambda\) going to infinity.
Analyzing the divergence structure of the self-energy of the halo propagator, we identified  two terms which are divergent in \(\Lambda\).
Therefore, two parameters which run with the cutoff are required, i.e., the coupling and the bare three-body binding energy.
The self-energy thus requires two renormalization conditions, one more than previously used.
In the simplest scenario either the charge or the matter radius can be used as a renormalization condition.
Once one of the two is known, the other radius can be predicted.
Moreover with one of the two radii also other observables which can be calculated in this theory can be predicted.

To check the range of applicability of the theory we compare the matter radius as function of the charge radius with predictions of standard Halo EFT, where in addition to the neutron-neutron interaction and the three-body interaction also a neutron-core interaction is present.
We calculated the charge and matter radii in Halo EFT for \(^6\)He, \(^{11}\)Li, \(^{14}\)Be, \(^{17}\)B, \(^{19}\)B, and \(^{22}\)C and compared to the matter radius as function of charge radius curves of {\color{black}the HS scheme}.
While only \(^6\)He and \(^{22}\)C are likely within the range of applicability of this theory, this comparison still gives useful insights.
We found that the relative deviations for the matter radii are not too large in all cases.
With 0.6~\% and 5.9~\% the smallest deviations are displayed by \(^6\)He and \(^{11}\)Li, while \(^{17}\)B has with 11.1~\% the largest deviation.
\textcolor{black}{It is important to note that the nuclei \(^{11}\)Li, \(^{14}\)Be, \(^{17}\)B, and \(^{19}\)B are outside of the naive region of applicability, i.e., the high momentum scale from the omission of the \(nc\) interaction is below the low momentum scale.
Therefore, these small deviations are quite remarkable and this requires further study.
For the radii, the HS scheme works much better than the power counting would suggest and is able to reproduce the more complex standard Halo EFT calculations quite well.}

Moreover, we compared also against the radii from Halo EFT calculations with artificially
increased or lowered neutron-core scattering lengths.
The smaller this scattering length, the larger the virtual-state energy becomes, and the deeper we move into the region of applicability.
For larger scattering lengths (artificially rescaled by factors of up to 10), we found that the data points in the matter-radius-over-charge-radius plot move away from the HS universal curves, as expected. Nevertheless, the relative deviations for the matter radii stayed below 13~\%, within uncertainty of the standard Halo EFT calculation.
When decreasing the scattering lengths, the data points move onto the universal curves, as expected.
At a rescaling of the scattering length by 0.0001 all data points where within the numerical uncertainties of the calculation on the universal prediction.
Thereby, we have confirmed that {\color{black}the HS scheme} is a special limit of the standard Halo EFT.
This finding complements the work by Naidon \cite{naidon} who has verified this limit behavior using different assumptions about the interactions.

It is worth noting that the running coupling of the trimer displays a Landau pole.
By using the charge or matter radii to renormalize the theory we were able to calculate the pole positions of the investigated nuclei.
While for \(^{14}\)Be and for \(^{17}\)B the Landau pole sets a high-momentum scale above
the high-momentum scale given by the core excitation, it is slightly lower than that scale in the case of \(^{11}\)Li and \(^{19}\)B.
However, the primary concern for the application to  these nuclei is the rather strong $nc$ interaction which 
leads to rather low "high-momentum" scales 
in the {\color{black}HS scheme}. In the case of \(^{6}\)He and  \(^{22}\)C, where this theory is a viable candidate, the Landau pole restricts the range of applicability significantly compared to the naive expectation. 
The rather good agreement for the radii in the comparison with standard Halo EFT including $nc$ interactions at leading order for all cases,
however, remains to be understood.

{\color{black} We conclude that the application of the HS scheme is favored compared to standard halo EFT when the neutron-core effective parameters are determined by a high-momentum scale of the theory. This is the case when there are no shallow bound or virtual states in the neutron-core system for the s-wave case. If neutron-core interactions in higher partial waves are strong (like the $^5$He resonance in the neutron-alpha interaction), it can still be treated as sub-leading for the low-energy properties of the neutron-neutron-core system. Since, the coupling $g_0^2$ in the HS scheme has a Landau pole, the breakdown scale of the theory cannot be estimated from an analysis of the scales \(M^\mathrm{nn}_\mathrm{hi}\) and \(M^\mathrm{nc}_\mathrm{hi}\) alone. An estimate of the breakdown scale thus always requires an analysis of the Landau pole, which might lie much lower as for the cases of $^6$He and $^{22}$C discussed above.}

Finally, we calculated 
the three-to-three scattering amplitude, and showed that it is properly renormalized.
As a cross check, we verified that it reproduces the correct imaginary part in the hypothetical case of a bound dineutron.
In the future, it might be interesting to apply this renormalized EFT also to other observables.
One could use again the charge radius for renormalization and then calculate the \(E1\) strength distribution of \(^{11}\)Li.
It could be compared to experimental data by Nakamura \textit{et al.} \cite{Nakamura:2006zz} as well as to data from full Halo EFT calculations \cite{Gobel:2022pvz}.
So far this comparison has been made only for the HS version of the EFT by estimating the trimer coupling \cite{Gobel:2022pvz}.

\begin{acknowledgments}
D.K. and H.W.H. acknowledge support by Deutsche Forschungsgemeinschaft (DFG, German Research Foundation) - Project-ID 279384907 - SFB 1245.
H.W.H. has been supported by the German Federal Ministry of Research, Technology and Space (BMFTR) (Grant No. 05P24RDB).
M.G. acknowledges support by the Czech Science Foundation GA\v{C}R Grant No.~25-18335S.
\end{acknowledgments}

\appendix

\section{Trimer Self-energy for $a<0$}\label{ap:self-energy}
We start from Eq.~\eqref{eq:halo_self_energy_first} to give a closed form expression for the integral $I_\Lambda(E)$ in the trimer self-energy. First we perform a contour integration for $q_0$ by inserting both propagators. Performing the integral transformation $\boldsymbol{q}\mapsto\boldsymbol{q}+\tfrac{m_\phi}{m_h}\boldsymbol{p}$ leads to
\begin{align}
\Sigma_h(p)&=-4\pi g_0^2\int\frac{\d^3q}{(2\pi)^3}f_a\left(-p_0+\frac{\boldsymbol{p}^2}{2m_h}+\frac{\boldsymbol{q}^2}{2\mu}-\i\eps\right)\,,\\
&=-4\pi g_0^2(\Lambda) I_\Lambda(-p_0+\tfrac{\boldsymbol{p}^2}{2m_h}-\i\eps)\,,
\end{align}
where $I_\Lambda(E)$ is defined as in Eq.~\eqref{eq:integral} and $\mu=2A/(A+2)$ is the reduced mass of the dimer-core system. Using the definition of the function $f_a$, Eq.~\eqref{eq:def_fa}, the integral reads
\begin{equation}\label{eq:integral_repr}
I_\Lambda(E)=\int_{q\leq\Lambda}\frac{\d^3q}{(2\pi)^3}\frac{1}{\sqrt{E+\tfrac{\boldsymbol{q}^2}{2\mu}}-\frac{1}{a}}=\frac{4\pi}{(2\pi)^3}\int_0^\Lambda\d q\frac{-aq^2}{\sqrt{a^2E+\tfrac{a^2}{2\mu}q^2}+1}\,,
\end{equation}
where we performed the trivial angular integration and factored out $a^{-1}<0$ from the denominator. Further appearing parameters are the UV cutoff $\Lambda>0$ and the reduced mass $\mu$. We simplify the integral using the substitution $y:=\tfrac{a^2}{2\mu}q^2$. This yields
\begin{equation}\label{eq:integral_representation}
I_\Lambda(E)=\frac{\mu^{3/2}}{\sqrt{2}\pi^2a^2}\int_0^{\frac{a^2\Lambda^2}{2\mu}}\d y\frac{\sqrt{y}}{\sqrt{a^2E+y}+1}\,.
\end{equation}
For this integral to be well-defined, we need the square root in the denominator to be evaluated in $E$ only on the domain $\mathbb{C}\setminus(-\infty,0]$ of complex numbers without the negative real numbers and without zero. There are no other restrictions on the integral since the integrand shows no poles. This is in fact a consequence of $a$ being negative as we explain below. Since $y$ starts from zero and then only attains positive values, we can restrict $E\in\mathbb{C}\setminus(-\infty,0]$. 

After having defined the integral, we now investigate its behavior for large $\Lambda$. In the integrand in Eq.~\eqref{eq:integral_representation} both numerator and denominator scale as $\sqrt{y}$ for large $y$. Therefore, the whole integral will scale as $I_\Lambda(E)\sim\Lambda^2$ when $\Lambda$ is taken large. But we make this statement more explicit by computing the integral and expanding it for large values of $\Lambda$. For this purpose we exploit Mathematica using the \texttt{Integrate} routine entering the definition in Eq.~\eqref{eq:integral_representation} as well as the aforementioned assumptions about our parameters. We present results\footnote{If we consider $\Im(E)=0$ when taking $E=B>0$ for our computations, we start again from the integral definition to get a closed form expression to avoid inconsistencies when continuing from complex to real $E$.} for $\Im(E)<0$. This yields
\begin{align}\nonumber
I_\Lambda(E)=
\frac{\mu^{3/2}}{2 \sqrt{2} \pi ^2 a^2}{\Bigg[}
&\frac{a^2 \Lambda  \sqrt{\Lambda ^2+2 \mu  E}}{\mu }+\frac{2 \sqrt{2} a \Lambda }{\sqrt{\mu }}+8 \sqrt{a^2 E-1} \arctan\left(\frac{1-a \sqrt{E}}{\sqrt{a^2 E-1}}\right)\\\nonumber
&-8 \sqrt{a^2 E-1} \arctan\left(\frac{\sqrt{2} a \left(\Lambda -\sqrt{\Lambda ^2+2 \mu  E}\right)+2 \sqrt{\mu }}{2 \sqrt{\mu  \left(a^2 E-1\right)}}\right)\\
&+(a^2E-2)\left(2\log(\sqrt{\Lambda^2+2\mu E}-\Lambda)-\log(2\mu E)\right)
{\Bigg]}\,.
\end{align}
This expression is the basis for computations in the main text where we need the exact numeric value of $I_\Lambda(E)$ or $I_\Lambda'(E)$. In some calculations we only need the leading terms. We found these terms by applying the \texttt{Series} routine to obtain a series of $I_\Lambda(E)$ in $\Lambda$ around $\Lambda=\infty$. We obtain
\begin{align}\nonumber
I_\Lambda(E)=&\frac{\Lambda^2\sqrt{\mu}}{2\sqrt{2}\pi^2}+\frac{\Lambda\mu}{\pi^2a}+\frac{\mu ^{3/2}}{2\sqrt{2} \pi ^2 a^2}{\Bigg[} (a^2E-2)\log \left(\frac{\mu  E}{2\Lambda ^2}\right)+a^2E\\\label{eq:integral_series}
&+8 \sqrt{a^2 E-1}\left(\arctan \left(\frac{1-a \sqrt{E}}{\sqrt{a^2 E-1}}\right)-\arccot(\sqrt{a^2E-1})\right){\Bigg]}+\mathcal{O}(\Lambda^{-1})\,.
\end{align}
Note that in the first two leading-order terms the coefficients do not depend on $E$. This implies that the difference $I_\Lambda(E_1)-I_\Lambda(E_2)$ behaves like $\ln(\Lambda)$ for large $\Lambda$.

Finally we want to compute the derivative $I_\Lambda'(E)$. This can be done straightforwardly using the explicit expressions above. Since the explicit expression for $I_\Lambda'(E)$ is rather long, we just state the leading terms in the series expansion. Again we employ Mathematica for this purpose:
\begin{align}\label{eq:integral_derivative_expr}
I_\Lambda'(E)=\frac{\mu^{3/2}}{2\sqrt{2}\pi^2} &{\Bigg[}\log\left(\frac{\mu E}{2\Lambda^2}\right)+2 
+\frac{4}{\sqrt{a^2E-1}}\left(\arctan \Bigg(\frac{1-a \sqrt{E}}{\sqrt{a^2 E-1}}\right)\nonumber \\
&\qquad -\arccot(\sqrt{a^2E-1})\Bigg){\Bigg]}+\mathcal{O}(\Lambda^{-1})\,.
\end{align}

\section{Comparison of Radii in Terms of Relative Deviations}
\label{ap:radii_rel_devs}

\begin{figure}[H]
    \centering
    \includegraphics[width=0.9\textwidth]{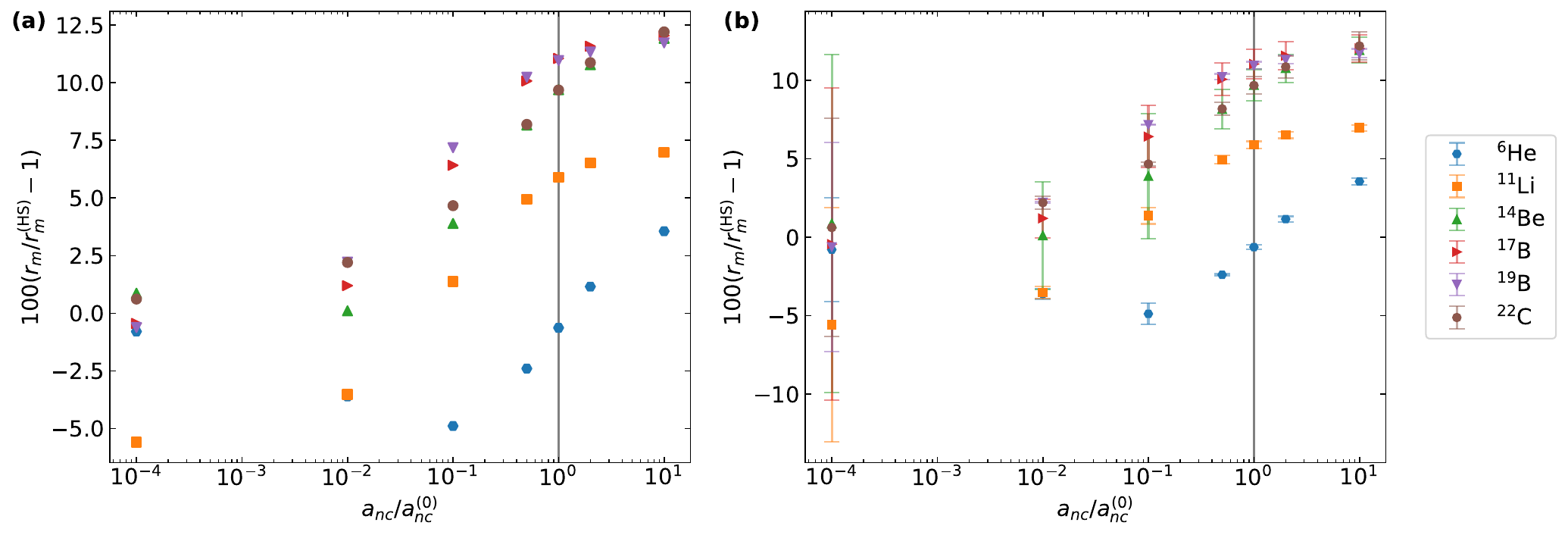}
    \caption{The relative deviations of the matter radii according the standard Halo EFT
    calculation from the prediction of {\color{black} the HS scheme} are shown in percent as function
    of the rescaling of the \(nc\) scattering length for different two neutron halo nuclei.
    While the left panel \textcolor{black}{(a)} shows the relative deviations themselves, the right panels \textcolor{black}{(b)} shows
    additionally also the uncertainty bars reflecting numerical and extraction uncertainties.}
    \label{fig:radii_rel_devs}
\end{figure}

In this appendix, we supplement on Fig.~\ref{fig:radii_calculation} by comparing the matter radii obtained with standard Halo EFT with the predictions of {\color{black}the HS scheme}.
We basically take the data points displayed in that figure representing the different nuclei at the different rescalings of the \(nc\) scattering length, calculate the HS matter radii by making use of their charge radius, and finally compare the actual matter radius to that prediction by calculating the relative deviation in percent.
The resulting plots can be found in Fig.~\ref{fig:radii_rel_devs}.
The values with error bars reflecting numerical and extraction uncertainties (see discussion in Appendix~\ref{ap:add_info_radii}) can be found in the right panel.
For clarity, in the left panel the same results are plotted without uncertainty bars.

One can nicely observe that making the \(nc\) scattering length smaller than its actual physical value brings the matter radii closer to the expectation based on {\color{black}the HS scheme}.
For most nuclei decreasing \(a_{nc}\) by a factor of 10,000 results in a relative deviation of the order of 1~\%.
The only exception is \(^{11}\)Li, where the deviation is larger.
Moreover, one can see that the relative deviation gets larger for increasing \(a_{nc}\) beyond its physical value.
However, even increasing the scattering length by a factor of 10 doesn't lead to a deviation larger than 12.5~\%.
For \(^{11}\)Li it even stays below 7.5~\% and for \(^6\)He below 5~\%.

Looking at the error bars in the right panel, we can conclude that for the rescaling of the scattering length by a factor of 10,000 within the error bars the matter radii of all nuclei are consistent with the HS prediction.
The error bars in this plot might appear large, but one has to keep in mind that these are the error bars for the relative deviations.
From Fig.~\ref{fig:heft_radii_wuncs} one can see that the uncertainties of the radii themselves are not large.
Moreover, even the relative errors of the radii are typically of the order or below 1~\%.
For the rescaling factor of 10,000 they are typically much larger, but still below 10~\%.
However, for the uncertainties of the relative deviation from the HS prediction that can result in quite some uncertainty.
As the prediction is based on the charge radius, this uncertainty is calculated from the uncertainty of matter radius in standard Halo EFT as well as from the uncertainty of the charge radius in standard Halo EFT.
Additionally, we note that for all nuclei except of \(^{11}\)Li and \(^6\)He within the error bars they converge from above against the limit of zero deviation for \(a_{nc} \to 0\).
In contrast to that we observe that \(^{11}\)Li is at a downscaling of \(a_{nc}\) by a factor of 100 within its entire error bar in the range of a negative relative deviation.
Similar behavior is observed for \(^6\)He.
This could be a specific property displayed by these nuclei. However, a shortcoming of the uncertainty estimation cannot be excluded.

\section{Additional Information on the Radii Obtained with Standard Halo EFT}
\label{ap:add_info_radii}

\begin{figure}[H]
    \centering
    \includegraphics[width=0.9\textwidth]{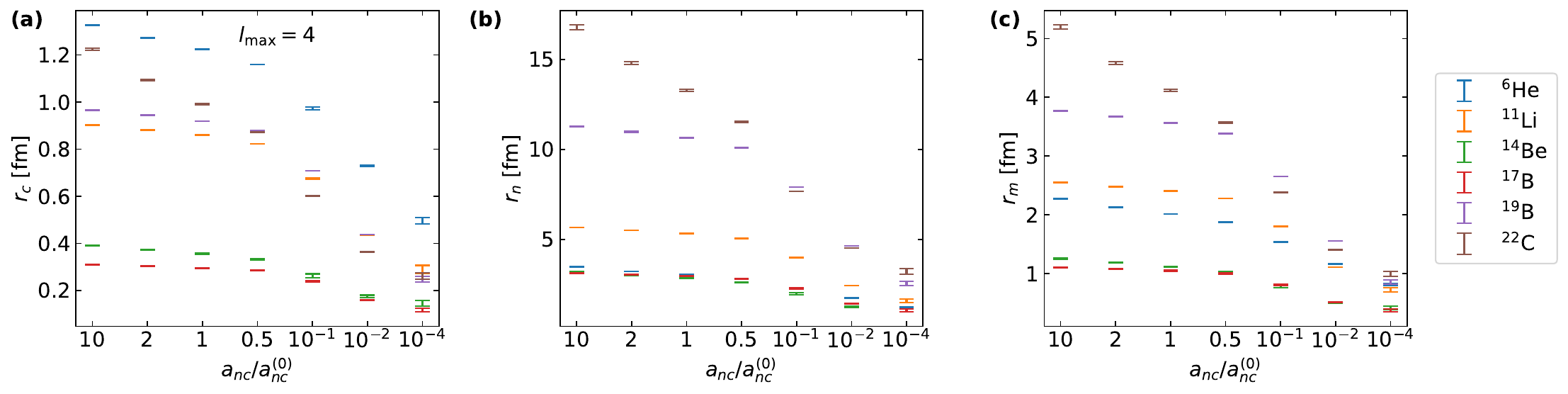}
    \caption{The radii of the different two-neutron halo nuclei obtained in standard Halo EFT are shown as a function of the employed \(nc\) scattering length over the physical one \(a_{nc}/a_{nc}^{(0)}\).
    The left panel \textcolor{black}{(a)} shows the radius \(r_c\), i.e., the average distance of the core from the overall
    center of mass, while the middle panel \textcolor{black}{(b)} shows \(r_n\), the average distance of the neutrons from the center of mass.
    The right panel \textcolor{black}{(c)} depicts the matter radius as function of the ratio of the \(nc\) scattering lengths.
    The bars give an uncertainty estimation taking into account the uncertainty of the extraction
    of the radius from the form factor as well as the numerical uncertainty.
    The numerical uncertainty was obtained by varying the number of mesh points by a factor of 2 and increasing the cutoff by 50~\%.
    }
    \label{fig:heft_radii_wuncs}
\end{figure}

In this appendix we provide additional information on the radii which have been calculated in standard Halo EFT.
Especially, we give another visualization of their dependence on the ratio of the actually used \(nc\) scattering length over the actual physical \(nc\) scattering length \(a_{nc}/a_{nc}^{(0)}\).
Moreover, we provide numerical and extraction uncertainty estimates.
The values of the radii as function of this ratio are shown in Fig.~\ref{fig:heft_radii_wuncs}.

The radii are extracted from the form factor by employing Eq.~\eqref{eq:y_i_deriv} and the conversions factors explained next to that equation.
The derivative of the form factor at \(k=0\) in \(k^2\) is calculated by the difference quotient based on the form factor values at \(k=10^{-3}\)~MeV and at \(k=4\)~MeV.
In order to estimate the extraction uncertainty the radii are also extracted using 5~MeV instead of 4~MeV for the second value.
Additionally, the numerical uncertainties are estimated by varying the cutoff \(\Lambda\) by 50~\% and the number of mesh points by a factor of 2.
The cutoff \(\Lambda\) is used as a three-body cutoff as well as to regulate the zero-range interactions, which come in momentum space with a Heaviside form factor.
The depicted central values are based on the higher cutoff and the higher number of mesh points.
The differences between the results based on the different cutoff, grid point, and extraction setting are used as uncertainty estimation.
These are applied symmetrically, so that the length of the error bar is twice the size.

For the calculations with \(a_{nc}/a_{nc}^{(0)}\) of 10, 2, 1, 0.5, or 0.1 we used cutoffs of 1~GeV and 1.5~GeV, while for the ratios of 0.01 and 0.0001 we employed cutoffs of 4~GeV and 6~GeV.
This is because as can be seen from the radii in these cases the spatial extension of the halos decreases implying a larger extension in momentum space.
Therefore, higher cutoffs have to be employed.
Independently of the described variation of the number of mesh points, these calculations employ also generally a higher number of mesh points.

Looking at the uncertainties we observe well converged values with the trend to small but larger uncertainties for the most extreme scaling of the \(nc\) scattering length by 0.0001.
Also for that rescaling the convergence is sufficient for these investigations.
For even better convergence, most likely even higher cutoffs instead of 6~GeV (and 4~GeV as setting to compare to) would be necessary.
For \(r_c\) we observe that across all these nuclei for all rescalings except the 0.0001 one the uncertainty is well below 0.01~fm.
For the rescaling of 0.0001 it is around 0.01~fm, but for all nuclei still smaller than 0.02~fm.
In the case of \(r_m\) we observe for all investigated systems except of \(^{22}\)C that the uncertainties for all rescaling settings except of 0.0001 are below 0.02~fm.
The results based on the scaling 0.0001 have uncertainties below 0.04~fm.
In the case of \(^{22}\)C we observe across all employed \(nc\) scattering lengths uncertainties below 0.05~fm.

Regarding the aspect of convergence, it is also worth mentioning that we evaluate Eqs.~(\ref{eq:F_i_2}) and~(\ref{eq:F_i_aavg}) in a partial wave basis.
In this basis the \(J\) and \(M\) quantum numbers are fixed due to the properties of the halo nucleus.
Moreover, the spins are constrained, the parity is fixed, and the rules of angular momentum algebra apply.
In order to arrive at finitely many states to be considered, we truncate the orbital quantum number of the \(ij\) subsystem \(l\) as well as the orbital quantum number of the movement of particle \(k\) relative to the \(ij\) subsystem \(\lambda\) by \(l_{\mathrm{max}}\).
Hereby the condition \(l \leq l_{\mathrm{max}} \, \land \, \lambda \leq l_{\mathrm{max}}\) applies.
We generally use \(l_{\mathrm{max}} = 4\).
However, for the considered \(2n\) halo nuclei with only s-wave interactions, the relative deviations between the \(l_{\mathrm{max}}=0\) and \(l_{\mathrm{max}}=4\) results is below 2~\% for \(r_c\) and below 3~\% for \(r_n\) and \(r_m\).
This holds across all the tested rescalings of the \(nc\) scattering length.

Finally, let us do a short comparison of our results at the physical \(nc\) scattering length with some literature values as a consistency check.
For that we use \(^{11}\)Li.
For \(r_c\) we have 0.86~fm or respectively 0.87~fm, if \(l_{\mathrm{max}}\) is zero instead of 4.
As expected this directly coincides with the Halo EFT calculation in Ref.~\cite{Gobel:2022pvz}, which has the same physics content with a limitation to \(l_{\mathrm{max}}=0\).
We can also compare with the Halo EFT calculations of Ref.~\cite{Canham:2008jd} using a \(nc\) virtual state energy of 25~keV and a two-neutron separation energy of 247~keV.
Given these differences, our value seems to agree with the result of that reference given by 1.0~fm.
For \(r_n\) they obtained 6.5~fm.
Our result of 5.3~fm (5.4~fm for \(l_{\mathrm{max}}=0\)) is compatible with that.
The three-body model calculations of Zhukov \textit{et al.} \cite{Zhukov:1993aw} yield similar ranges for the radii.

\section{Trimer Self-energy and Imaginary Part for $a>0$}\label{ap:imag_part_check}
To check the consistency of the halo propagator with the well-known effective-range expansion, we consider dimer-core scattering. Since it applies to a system of two particles, we must modify the theory such that the dineutron in our description corresponds to a bound particle. This is settled by taking the scattering length of the two identical particles to be positive, $ a>0$. Although the dressed dimer propagator does not change its form under this sign change, one must keep track of how this sign change affects the previous results. In fact a new pole is added in Eq.~\eqref{eq:halo_self_energy_first} if $a>0$. Fortunately, this pole is not enclosed by the contour that we chose in this computation. But one physical change occurs, which is that $B>1/a^2$ holds. Otherwise, the system would be unstable against dimer formation and core emission. Thus, the inverse scattering amplitude for the dimer-core pair at rest and $-1/a^2<p_0<0$, where a bound dimer exists, is given by
\begin{equation}
    t^{-1}_\mathrm{dc}(p_0)=(Z_dg_0^2G_h(p_0,\boldsymbol{0}))^{-1}=\frac{a}{8\pi g_0^2}(p_0+B)+\frac{a}{2}(I_\Lambda(-p_0-\i\eps)-I_\Lambda(B))\,,
\end{equation}
where $Z_d=8\pi/a$ is the field strength renormalization of the dimer field.
The effective-range expansion provides a consistency check by the unitarity term, i.e., every two-body t-matrix must satisfy $-\tfrac{2\pi}{\mu}\Im\{t^{-1}(E)\}=-\sqrt{2\mu E}$. In our case $E=1/a^2+p_0$ is the energy above the threshold for a bound dimer. The imaginary part of the inverse scattering amplitude is in our case given by
\begin{equation}
    \Im\{t^{-1}_\mathrm{dc}(p_0)\}=\frac{a}{2}\Im\{I_\Lambda(-p_0-\i\eps)\}\,.
\end{equation}
The evaluation of this imaginary part involves the computation of the integral in terms of analytical expressions. For $a>0$ the first equality of Eq.~\eqref{eq:integral_repr} is still valid, but now the integrand exhibits a pole for certain values of $E$. We show this by doing the same substitution $y:=\tfrac{a^2}{2\mu}q^2$ as before and keeping in mind that $a>0$. We arrive at
\begin{equation}\label{eq:integral_representation_a}
I_\Lambda(E)=\frac{\mu^{3/2}}{\sqrt{2}\pi^2a^2}\int_0^{\frac{a^2\Lambda^2}{2\mu}}\d y\frac{\sqrt{y}}{\sqrt{a^2E+y}-1}\,.
\end{equation}
To avoid the pole, the square root in the denominator must be greater than 1, if it is not complex-valued. This corresponds to $E>1/a^2$. Thus, the domain for $E$ is the set $\mathbb{C}\setminus(-\infty,1/a^2]$. In order to obtain the imaginary part of $I_\Lambda(-p_0-\i\eps)$ where $0>p_0>-1/a^2$, we follow the same evaluation steps as in the last section in the appendix. We do not state the result of the integral, but directly write down the terms one gets from the evaluation of the imaginary part using Mathematica\footnote{One might need to transform trigonometric functions to exponential functions, in this case logarithms.}:
\begin{align}\nonumber
\Im\{I_\Lambda(&-p_0-\i\eps)\}=
\frac{\mu^{3/2}}{2\sqrt{2}a^2\pi^2}
\Im{\Bigg\{}4 X_1 \log \left(2-\sqrt{\frac{2}{\mu}}X_2\right)+\frac{a \Lambda  X_2}{\mu }\\\nonumber
&+2 \left(X_1^2-1 \right) \log \left(\Lambda +\frac{X_2}{a}\right)\\\nonumber
&-4 X_1 \log \left(\sqrt{2} \left(\Lambda  X_1+\frac{X_2}{\mu}\right)+2 a \sqrt{\mu } p_0+2\i a \sqrt{\mu } \eps \right)\\\nonumber
&+\log (2\mu) \left(1-X_1^2\right)+2 \log (\mu ) X_1\\\nonumber
&-4 X_1 \log \left(1-a \sqrt{-p_0-\i \eps }\right)+4 X_1 \log \left(a p_0+\i a \eps +\sqrt{-p_0-\i \eps }\right)\\\label{eq:imag_part_looong}
&-\left(X_1^2+1\right) \log \left(-p_0-\i \eps \right)+4 \log \left(\Lambda +\frac{X_2}{a}\right){\Bigg\}}\,,
\end{align}
where we abbreviated $X_1=\sqrt{a^2 p_0+\i a^2 \eps +1}$ and $X_2=\sqrt{a^2\Lambda ^2-2 \mu  a^2p_0-2\i \mu a^2 \eps}$. Out of these ten terms, there is only one term that contributes to the imaginary part. For the other nine terms one needs to assume $\Lambda$ to be sufficiently large, but finite, in order to get some square roots real. Using $0>p_0>-1/a^2$ and taking the limit $\eps\to0$ afterwards, one can find that these other nine terms do not contribute to the imaginary part. The only term left is the one in the first line of Eq.~\eqref{eq:imag_part_looong}:
\begin{align}
\lim_{\eps\to0}\Im\{I_\Lambda(-p_0-\i\eps)\}&=\frac{2\mu^{3/2}}{\sqrt{2}a^2\pi^2}\lim_{\eps\to0}\Im\left(X_1 \log \left(2-\sqrt{\frac{2}{\mu }}X_2\right)\right\}\\
&=\frac{2\mu^{3/2}}{\sqrt{2}a\pi^2}\lim_{\eps\to0}\sqrt{p_0 +\frac{1}{a^2}+\i \eps}\,\,\Im\left\{\log \left(2-\sqrt{\frac{2}{\mu }}X_2\right)\right\}\\
&=\frac{2\mu^{3/2}}{\sqrt{2}a\pi}\sqrt{p_0 +\frac{1}{a^2}}\,.
\end{align}
The square root $X_1=\sqrt{a^2 p_0+\i a^2 \eps +1}$ can be factored out since it is a positive number after taking $\eps\to0$ in the end. The remaining imaginary part of the logarithm can be calculated to be $\pi$ when $\Lambda$ is sufficiently large and after taking $\eps\to0$ in the end. We verified this last statement using a sufficiently large, but finite value of $\Lambda$ and evaluated the limit in Mathematica. Altogether we have
\begin{equation}
    -\frac{2\pi}{\mu}\Im\{t^{-1}_\mathrm{dc}(p_0)\}=-\sqrt{2\mu\left(p_0 +\frac{1}{a^2}\right)}\,.
\end{equation}

\section{Scattering Amplitude and Renormalization}\label{ap:reno_scat_amp}
In this appendix we show that the result for the neutron-neutron-core scattering amplitudes behaves well with respect to the cutoff $\Lambda$. Moreover, we derive how $g_0^{-2}$ depends on $\Lambda$ when a scattering amplitude is used as a second renormalization condition.

First, we derive an expression for $(g_0^2G_h(\widetilde{E},\boldsymbol{0}))^{-1}$, which is the not fully evaluated part in the scattering amplitude in Eq.~\eqref{eq:scat_amp_nnc}. For that purpose, we start with Eq.~\eqref{eq:G_h_without_B0}, which in our case evaluates to
\begin{equation}
	G_h^{-1}(\widetilde{E},\boldsymbol{0})=\widetilde{E}+B-4\pi g_0^2I_\Lambda(B)+4\pi g_0^2I_\Lambda(-\widetilde{E}-\i\eps)\,.
\end{equation}
Multiplying with $g_0^{-2}$ and using the expression given in Eq.~\eqref{eq:g0-2}, we arrive at
\begin{align}\label{eq:g02Gh-1}
	\left(g_0^2G_h(\widetilde{E},\boldsymbol{0})\right)^{-1}&=g_0^{-2}(\Lambda)(\widetilde{E}+B)+[4\pi I_\Lambda(-\widetilde{E}-\i\eps)-I_\Lambda(B)]\\
	&=\left[\frac{K_\mathrm{c/m}}{\E{r_\mathrm{c/m}^2}_\mathrm{exp}}+4\pi I_\Lambda'(B)\right](\widetilde{E}+B)+[4\pi I_\Lambda(-\widetilde{E}-\i\eps)-I_\Lambda(B)]\\
	&=\frac{K_\mathrm{c/m}}{\E{r_\mathrm{c/m}^2}_\mathrm{exp}}(\widetilde{E}+B)+4\pi\left[(\widetilde{E}+B)I_\Lambda'(B)+I_\Lambda(-\widetilde{E}-\i\eps)-I_\Lambda(B)\right]\,,
\end{align}
where in the last line all the $\Lambda$-dependent terms are summarized in the square brackets. To check whether these terms depend on $\Lambda$ only at some suppressed order $\Lambda^{-1}$, we use our explicit expressions for the integrals. Using the Mathematica routine \texttt{Series} to get a series in $\Lambda$ around $\Lambda=\infty$, we get a rather long, but explicit expression. The leading term of the series is independent of $\Lambda$ and the corrections are of order $\Lambda^{-2}$. Thus, one can show that our expression for the scattering amplitude behaves well with respect to the cutoff.

Our second derivation in this appendix concerns the $\Lambda$-dependence of $g_0$, when a scattering amplitude is used as input for a second renormalization condition. Inserting Eq.~\eqref{eq:g02Gh-1} into Eq.~\eqref{eq:scat_amp_nnc} we arrive at Eq.~\eqref{eq:f_B0_eliminated}, which determines $g_0^{-2}(\Lambda)$. Solving for $g_0^{-2}(\Lambda)$ yields
\begin{equation}
	\frac{1}{g_0^2(\Lambda)}=\frac{1}{\widetilde{E}+B}\left(\frac{d}{t_\mathrm{exp}}-4\pi\left[I_\Lambda(-\widetilde{E}-\i\eps)-I_\Lambda(B)\right]\right)\,.
\end{equation}
Again, the $\Lambda$-dependent terms are summarized in the square brackets. Finally, we apply the routine \texttt{Series} from Mathematica expanding these terms at $\Lambda=\infty$. As before, the result is a large set of terms, but the leading order dependence is $\ln(\Lambda)$, for which we state the coefficient:
\begin{equation}
	I_\Lambda(-\widetilde{E}-\i\eps)-I_\Lambda(B)=\frac{\mu^{3/2}}{\sqrt{2}\pi^2}(B+\widetilde{E})\ln(\Lambda)+\const+\mathcal{O}(\Lambda^{-1})\,.
\end{equation}
For scattering we need the on-shell t-matrix elements having $\widetilde{E}>0$ and hence the term $-4\pi[I_\Lambda(-\widetilde{E}-\i\eps)-I_\Lambda(B)]$ tends to $-\infty$ when $\Lambda$ becomes large. Both things were shown in this case, the dependence on $\Lambda$ as well as the existence of a Landau pole.


\bibliography{bibliography.bib}

@article{Ryberg:2017tpv,
    author = "Ryberg, Emil and Forss{\'e}n, Christian and Platter, Lucas",
    title = "{Three-Body Halo States in Effective Field Theory: Renormalization and Three-Body Interactions in the Helium-6 System}",
    eprint = "1701.08576",
    archivePrefix = "arXiv",
    primaryClass = "nucl-th",
    doi = "10.1007/s00601-017-1307-1",
    journal = "Few Body Syst.",
    volume = "58",
    number = "4",
    pages = "143",
    year = "2017"
}

@article{Rotureau:2012yu,
    author = "Rotureau, J. and van Kolck, U.",
    title = "{Effective Field Theory and the Gamow Shell Model: The $^6He$ Halo Nucleus}",
    eprint = "1201.3351",
    archivePrefix = "arXiv",
    primaryClass = "nucl-th",
    doi = "10.1007/s00601-012-0455-6",
    journal = "Few Body Syst.",
    volume = "54",
    pages = "725--735",
    year = "2013"
}

@article{Bedaque:2002yg,
    author = "Bedaque, Paulo F. and Rupak, Gautam and Griesshammer, Harald W. and Hammer, Hans-Werner",
    title = "{Low-energy expansion in the three-body system to all orders and the triton channel}",
    eprint = "nucl-th/0207034",
    archivePrefix = "arXiv",
    reportNumber = "LBL-50137, TUM-T39-02-12",
    doi = "10.1016/S0375-9474(02)01402-1",
    journal = "Nucl. Phys. A",
    volume = "714",
    pages = "589--610",
    year = "2003"
}

@article{Hammer:2017tjm,
    author = "Hammer, H. -W. and Ji, C. and Phillips, D. R.",
    title = "{Effective field theory description of halo nuclei}",
    eprint = "1702.08605",
    archivePrefix = "arXiv",
    primaryClass = "nucl-th",
    doi = "10.1088/1361-6471/aa83db",
    journal = "J. Phys. G",
    volume = "44",
    number = "10",
    pages = "103002",
    year = "2017"
}

@article{Hammer:2019poc,
    author = {Hammer, H. -W. and K{\"o}nig, S. and van Kolck, U.},
    title = "{Nuclear effective field theory: status and perspectives}",
    eprint = "1906.12122",
    archivePrefix = "arXiv",
    primaryClass = "nucl-th",
    doi = "10.1103/RevModPhys.92.025004",
    journal = "Rev. Mod. Phys.",
    volume = "92",
    number = "2",
    pages = "025004",
    year = "2020"
}

@article{fr_heft,
   author = "Göbel, Matthias and Hammer, Hans-Werner and Phillips, Daniel R.",
   journal = "in preparation",
   year = "2026"
}

@article{GonzalezTrotter:1999zz,
    author = "Gonzalez Trotter, D. E. and others",
    title = "{New Measurement of the S-10 Neutron-Neutron Scattering Length Using the Neutron-Proton Scattering Length as a Standard}",
    doi = "10.1103/PhysRevLett.83.3788",
    journal = "Phys. Rev. Lett.",
    volume = "83",
    pages = "3788--3791",
    year = "1999"
}

@article{Costa:2025ldi,
    author = "Costa, Davi B. and Hongo, Masaru and Son, Dam Thanh",
    title = "{Effective field theory for weakly bound two-neutron halo nuclei: Corrections from neutron-neutron effective range}",
    eprint = "2503.18519",
    archivePrefix = "arXiv",
    primaryClass = "nucl-th",
    reportNumber = "RIKEN-iTHEMS-Report-25",
    doi = "10.1103/lds3-g3tp",
    journal = "Phys. Rev. C",
    volume = "112",
    number = "1",
    pages = "014001",
    year = "2025"
}

@article{naidon,
    author = "Naidon, Pascal",
    title = "{Universal geometry of two-neutron halos and Borromean Efimov states close to dissociation}",
    eprint = "2302.08716",
    archivePrefix = "arXiv",
    primaryClass = "nucl-th",
    doi = "10.21468/SciPostPhys.15.3.123",
    journal = "SciPost Phys.",
    volume = "15",
    number = "3",
    pages = "123",
    year = "2023"
}

@article{Fahlander_2013,
doi = {10.1088/0031-8949/2013/T152/010301},
url = {https://dx.doi.org/10.1088/0031-8949/2013/T152/010301},
year = {2013},
month = {jan},
publisher = {},
volume = {2013},
number = {T152},
pages = {010301},
author = {Fahlander, Claes and Jonson, Björn},
title = {Nobel Symposium 152: Physics with Radioactive Beams},
journal = {Physica Scripta}
}

@Article{Jensen:2004zz,
  Title                    = {{Structure and reactions of quantum halos}},
  Author                   = {Jensen, A. S. and Riisager, K. and Fedorov, D. V. and
 Garrido, E.},
  Journal                  = {Rev. Mod. Phys.},
  Year                     = {2004},
  Pages                    = {215-261},
  Volume                   = {76},

  Doi                      = {10.1103/RevModPhys.76.215}
}

@Article{Riisager:2012it,
  Title                    = {{Halos and related structures}},
  Author                   = {Riisager, K.},
  Journal                  = {Phys. Scripta},
  Year                     = {2013},
  Pages                    = {014001},
  Volume                   = {T152},

  Doi                      = {10.1088/0031-8949/2013/T152/014001}
}

@Article{Jonson:2004,
  Title                    = {Light dripline nuclei},
  Author                   = {Jonson, B.},
  Journal                  = {Physics Reports},
  Year                     = {2004},
  Number                   = {1},
  Pages                    = {1 - 59},
  Volume                   = {389},

  Doi                      = {http://dx.doi.org/10.1016/j.physrep.2003.07.004},
  ISSN                     = {0370-1573}
}

@Article{Hansen:1995pu,
  Title                    = {{Nuclear halos}},
  Author                   = {Hansen, P. G. and Jensen, A. S. and Jonson, B.},
  Journal                  = {Ann. Rev. Nucl. Part. Sci.},
  Year                     = {1995},
  Pages                    = {591-634},
  Volume                   = {45},

  Doi                      = {10.1146/annurev.ns.45.120195.003111}
}

@article{Bertulani:2002sz,
    author = "Bertulani, C. A. and Hammer, H. W. and Van Kolck, U.",
    title = "{Effective field theory for halo nuclei}",
    eprint = "nucl-th/0205063",
    archivePrefix = "arXiv",
    doi = "10.1016/S0375-9474(02)01270-8",
    journal = "Nucl. Phys. A",
    volume = "712",
    pages = "37--58",
    year = "2002"
}

@article{Bedaque:2003wa,
    author = "Bedaque, P. F. and Hammer, H. W. and van Kolck, U.",
    title = "{Narrow resonances in effective field theory}",
    eprint = "nucl-th/0304007",
    archivePrefix = "arXiv",
    doi = "10.1016/j.physletb.2003.07.049",
    journal = "Phys. Lett. B",
    volume = "569",
    pages = "159--167",
    year = "2003"
}

@article{Canham:2008jd,
    author = "Canham, David L. and Hammer, H. -W.",
    title = "{Universal properties and structure of halo nuclei}",
    eprint = "0807.3258",
    archivePrefix = "arXiv",
    primaryClass = "nucl-th",
    reportNumber = "HISKP-TH-08-12",
    doi = "10.1140/epja/i2008-10632-4",
    journal = "Eur. Phys. J. A",
    volume = "37",
    pages = "367--380",
    year = "2008"
}

@article{Hammer:2011ye,
    author = "Hammer, H. -W. and Phillips, D. R.",
    title = "{Electric properties of the Beryllium-11 system in Halo EFT}",
    eprint = "1103.1087",
    archivePrefix = "arXiv",
    primaryClass = "nucl-th",
    reportNumber = "INT-PUB-11-007",
    doi = "10.1016/j.nuclphysa.2011.06.028",
    journal = "Nucl. Phys. A",
    volume = "865",
    pages = "17--42",
    year = "2011"
}

@article{Elkamhawy:2019nxq,
    author = "Elkamhawy, Wael and Yang, Zichao and Hammer, Hans-Werner and Platter, Lucas",
    title = "{$\mathbf{\beta}$-delayed proton emission from $\mathbf{^{11}}$Be in effective field theory}",
    eprint = "1909.12206",
    archivePrefix = "arXiv",
    primaryClass = "nucl-th",
    reportNumber = "INT-PUB-19-045",
    doi = "10.1016/j.physletb.2021.136610",
    journal = "Phys. Lett. B",
    volume = "821",
    pages = "136610",
    year = "2021"
}

@article{Gobel:2021pvw,
    author = {G\"obel, Matthias and Aumann, Thomas and Bertulani, Carlos A. and Frederico, Tobias and Hammer, Hans-Werner and Phillips, Daniel R.},
    title = "{Neutron-neutron scattering length from the He6(p,p\ensuremath{\alpha})nn reaction}",
    eprint = "2103.03224",
    archivePrefix = "arXiv",
    primaryClass = "nucl-th",
    doi = "10.1103/PhysRevC.104.024001",
    journal = "Phys. Rev. C",
    volume = "104",
    number = "2",
    pages = "024001",
    year = "2021"
}

@article{Gobel:2022pvz,
    author = {G\"obel, Matthias and Acharya, Bijaya and Hammer, Hans-Werner and Phillips, Daniel R.},
    title = "{Final-state interactions and spin structure in E1 breakup of Li11 in halo effective field theory}",
    eprint = "2207.14281",
    archivePrefix = "arXiv",
    primaryClass = "nucl-th",
    doi = "10.1103/PhysRevC.107.014617",
    journal = "Phys. Rev. C",
    volume = "107",
    number = "1",
    pages = "014617",
    year = "2023"
}

@article{Tanihata:1985psr,
    author = "Tanihata, I. and Hamagaki, H. and Hashimoto, O. and Shida, Y. and Yoshikawa, N. and Sugimoto, K. and Yamakawa, O. and Kobayashi, T. and Takahashi, N.",
    title = "{Measurements of Interaction Cross-Sections and Nuclear Radii in the Light p-Shell Region}",
    doi = "10.1103/PhysRevLett.55.2676",
    journal = "Phys. Rev. Lett.",
    volume = "55",
    pages = "2676--2679",
    year = "1985"
}

@article{Hansen:1987mc,
    author = "Hansen, P. G. and Jonson, B.",
    title = "{The Neutron halo of extremely neutron-rich nuclei}",
    doi = "10.1209/0295-5075/4/4/005",
    journal = "EPL",
    volume = "4",
    pages = "409--414",
    year = "1987"
}

@article{HongoSon22,
	title = {Universal Properties of Weakly Bound Two-Neutron Halo Nuclei},
	author = {Hongo, Masaru and Son, Dam Thanh},
	journal = {Phys. Rev. Lett.},
	volume = {128},
	issue = {21},
	pages = {212501},
	numpages = {6},
	year = {2022},
	month = {May},
	publisher = {American Physical Society},
	doi = {10.1103/PhysRevLett.128.212501},
  	url = {https://link.aps.org/doi/10.1103/PhysRevLett.128.212501}
}

@article{tanaka_c22,
	title = {Observation of a Large Reaction Cross Section in the Drip-Line Nucleus $^{22}${C}},
	author = {Tanaka, K. and Yamaguchi, T. and Suzuki, T. and Ohtsubo, T. and Fukuda, M. and Nishimura, D. and Takechi, M. and Ogata, K. and Ozawa, A. and Izumikawa, T. and Aiba, T. and Aoi, N. and Baba, H. and Hashizume, Y. and Inafuku, K. and Iwasa, N. and Kobayashi, K. and Komuro, M. and Kondo, Y. and Kubo, T. and Kurokawa, M. and Matsuyama, T. and Michimasa, S. and Motobayashi, T. and Nakabayashi, T. and Nakajima, S. and Nakamura, T. and Sakurai, H. and Shinoda, R. and Shinohara, M. and Suzuki, H. and Takeshita, E. and Takeuchi, S. and Togano, Y. and Yamada, K. and Yasuno, T. and Yoshitake, M.},
	journal = {Phys. Rev. Lett.},
	volume = {104},
	issue = {6},
	pages = {062701},
	numpages = {4},
	year = {2010},
	month = {Feb},
	publisher = {American Physical Society},
	doi = {10.1103/PhysRevLett.104.062701},
	url = {https://link.aps.org/doi/10.1103/PhysRevLett.104.062701}
}

@article{acharya_c22,
	title = {Implications of a matter-radius measurement for the structure of Carbon-22},
	journal = {Physics Letters B},
	volume = {723},
	number = {1},
	pages = {196-200},
	year = {2013},
	issn = {0370-2693},
	doi = {https://doi.org/10.1016/j.physletb.2013.04.055},
	url = {https://www.sciencedirect.com/science/article/pii/S0370269313003444},
	author = {B. Acharya and C. Ji and D.R. Phillips}
}

@article{mosby_c22,
	title = {Search for $^{21}${C} and constraints on $^{22}${C}},
	journal = {Nuclear Physics A},
	volume = {909},
	pages = {69-78},
	year = {2013},
	issn = {0375-9474},
	doi = {https://doi.org/10.1016/j.nuclphysa.2013.04.004},
	url = {https://www.sciencedirect.com/science/article/pii/S0375947413004843},
	author = {S. Mosby and N.S. Badger and T. Baumann and D. Bazin and M. Bennett and J. Brown and G. Christian and P.A. DeYoung and J.E. Finck and M. Gardner and J.D. Hinnefeld and E.A. Hook and E.M. Lunderberg and B. Luther and D.A. Meyer and M. Mosby and G.F. Peaslee and W.F. Rogers and J.K. Smith and J. Snyder and A. Spyrou and M.J. Strongman and M. Thoennessen}
}

@misc{levels_c20,
	note = {From ENSDF database as of June 9, 2017}
}

@article{ex_state_he4,
    title = {Energy levels of light nuclei {A} = 4},
    journal = {Nuclear Physics A},
    volume = {541},
    number = {1},
    pages = {1-104},
    year = {1992},
    issn = {0375-9474},
    doi = {https://doi.org/10.1016/0375-9474(92)90635-W},
    url = {https://www.sciencedirect.com/science/article/pii/037594749290635W},
    author = {D.R. Tilley and H.R. Weller and G.M. Hale}
}

@article{ex_states_li9,
    title = {Energy levels of light nuclei {A}=8,9,10},
    journal = {Nuclear Physics A},
    volume = {745},
    number = {3},
    pages = {155-362},
    year = {2004},
    issn = {0375-9474},
    doi = {https://doi.org/10.1016/j.nuclphysa.2004.09.059},
    url = {https://www.sciencedirect.com/science/article/pii/S0375947404010267},
    author = {D.R. Tilley and J.H. Kelley and J.L. Godwin and D.J. Millener and J.E. Purcell and C.G. Sheu and H.R. Weller}
}

@article{ex_states_be12,
    title = {Energy levels of light nuclei {A}=12},
    journal = {Nuclear Physics A},
    volume = {968},
    pages = {71-253},
    year = {2017},
    issn = {0375-9474},
    doi = {https://doi.org/10.1016/j.nuclphysa.2017.07.015},
    url = {https://www.sciencedirect.com/science/article/pii/S0375947417303330},
    author = {J.H. Kelley and J.E. Purcell and C.G. Sheu}
}

@article{ex_states_b15,
    author    = {M. Stanoiu and M. Belleguic and Z. Dombr{\'a}di and others},
    title     = {Observation of bound excited states in $^{15}${B}},
    journal   = {The European Physical Journal A},
    volume    = {22},
    number    = {1},
    pages     = {5--8},
    year      = {2004},
    publisher = {Springer},
    doi = {https://doi.org/10.1140/epja/i2004-10078-8}
}

@article{ex_states_b17,
    title = {Excited states in neutron rich boron isotopes},
    journal = {Physics Letters B},
    volume = {608},
    number = {3},
    pages = {206-214},
    year = {2005},
    issn = {0370-2693},
    doi = {https://doi.org/10.1016/j.physletb.2005.01.022},
    url = {https://www.sciencedirect.com/science/article/pii/S0370269305000390},
    author = {R. Kanungo and Z. Elekes and H. Baba and Zs. Dombr{\'a}di and Zs. F{\"u}l{\"o}p and J. Gibelin and {\'A}. Horv{\'a}th and Y. Ichikawa and E. Ideguchi and N. Iwasa and H. Iwasaki and S. Kawai and Y. Kondo and T. Motobayashi and M. Notani and T. Ohnishi and A. Ozawa and H. Sakurai and S. Shimoura and E. Takeshita and S. Takeuchi and I. Tanihata and Y. Togano and C. Wu and Y. Yamaguchi and Y. Yanagisawa and A. Yoshida and K. Yoshida}
}

@article{braaten_dimer,
	title = {Universality in few-body systems with large scattering length},
	journal = {Physics Reports},
	volume = {428},
	number = {5},
	pages = {259-390},
	year = {2006},
	issn = {0370-1573},
	doi = {https://doi.org/10.1016/j.physrep.2006.03.001},
	url = {https://www.sciencedirect.com/science/article/pii/S0370157306000822},
	author = {Eric Braaten and H.-W. Hammer}
}

@article{togano:2016,
        title = {Interaction cross section study of the two-neutron halo nucleus 22{C}},
        journal = {Physics Letters B},
        volume = {761},
        pages = {412-418},
        year = {2016},
        issn = {0370-2693},
        doi = {https://doi.org/10.1016/j.physletb.2016.08.062},
        url = {https://www.sciencedirect.com/science/article/pii/S0370269316304890},
        author = {Y. Togano and T. Nakamura and Y. Kondo and J.A. Tostevin and A.T. Saito and J. Gibelin and N.A. Orr and N.L. Achouri and T. Aumann and H. Baba and F. Delaunay and P. Doornenbal and N. Fukuda and J.W. Hwang and N. Inabe and T. Isobe and D. Kameda and D. Kanno and S. Kim and N. Kobayashi and T. Kobayashi and T. Kubo and S. Leblond and J. Lee and F.M. Marqués and R. Minakata and T. Motobayashi and D. Murai and T. Murakami and K. Muto and T. Nakashima and N. Nakatsuka and A. Navin and S. Nishi and S. Ogoshi and H. Otsu and H. Sato and Y. Satou and Y. Shimizu and H. Suzuki and K. Takahashi and H. Takeda and S. Takeuchi and R. Tanaka and A.G. Tuff and M. Vandebrouck and K. Yoneda}
}

@article{Gobel:2019jba,
    author = {G{\"o}bel, M. and Hammer, H. -W. and Ji, C. and Phillips, D. R.},
    title = "{Momentum-Space Probability Density of $^6$He in Halo Effective Field Theory}",
    eprint = "1904.07182",
    archivePrefix = "arXiv",
    primaryClass = "nucl-th",
    doi = "10.1007/s00601-019-1528-6",
    journal = "Few Body Syst.",
    volume = "60",
    number = "4",
    pages = "61",
    year = "2019"
}

@article{Kirchner:2024axc,
    author = {Kirchner, Tanja and G{\"o}bel, Matthias and Hammer, Hans-Werner},
    title = "{Neutron-neutron distribution of the triton from pionless effective field theory}",
    eprint = "2410.13789",
    archivePrefix = "arXiv",
    primaryClass = "nucl-th",
    doi = "10.1103/PhysRevC.111.044002",
    journal = "Phys. Rev. C",
    volume = "111",
    number = "4",
    pages = "044002",
    year = "2025"
}

@phdthesis{Gobel:2024ovk,
    author = {G{\"o}bel, Matthias},
    title = "{Structure and breakup reactions of neutron halo nuclei}",
    doi = "10.26083/tuprints-00027581",
    school = "Darmstadt, Tech. U.",
    year = "2024"
}

@article{Wang:2021xhn,
    author = "Wang, Meng and Huang, W. J. and Kondev, F. G. and Audi, G. and Naimi, S.",
    title = "{The AME 2020 atomic mass evaluation (II). Tables, graphs and references}",
    doi = "10.1088/1674-1137/abddaf",
    journal = "Chin. Phys. C",
    volume = "45",
    number = "3",
    pages = "030003",
    year = "2021"
}

@article{Zhukov:1993aw,
    author = "Zhukov, M. V. and Danilin, B. V. and Fedorov, D. V. and Bang, J. M. and Thompson, I. J. and Vaagen, J. S.",
    title = "{Bound state properties of Borromean Halo nuclei: He-6 and Li-11}",
    reportNumber = "NORDITA-92-90-N",
    doi = "10.1016/0370-1573(93)90141-Y",
    journal = "Phys. Rept.",
    volume = "231",
    pages = "151--199",
    year = "1993"
}

@article{Nakamura:2006zz,
    author = "Nakamura, T. and others",
    title = "{Observation of Strong Low-Lying E-1 Strength in the Two-Neutron Halo Nucleus Li-11}",
    doi = "10.1103/PhysRevLett.96.252502",
    journal = "Phys. Rev. Lett.",
    volume = "96",
    pages = "252502",
    year = "2006"
}

@article{Brodeur:2011sam,
    author = "Brodeur, M. and others",
    title = "{First Direct Mass Measurement of the Two-Neutron Halo Nucleus $^6$He and Improved Mass for the Four-Neutron Halo $^{8}$He}",
    eprint = "1107.1684",
    archivePrefix = "arXiv",
    primaryClass = "nucl-ex",
    doi = "10.1103/PhysRevLett.108.052504",
    journal = "Phys. Rev. Lett.",
    volume = "108",
    pages = "052504",
    year = "2012"
}

@article{Ji:2014wta,
    author = "Ji, C. and Elster, Ch. and Phillips, D. R.",
    title = "{$^6$He nucleus in halo effective field theory}",
    eprint = "1405.2394",
    archivePrefix = "arXiv",
    primaryClass = "nucl-th",
    reportNumber = "INT-PUB-14-012",
    doi = "10.1103/PhysRevC.90.044004",
    journal = "Phys. Rev. C",
    volume = "90",
    number = "4",
    pages = "044004",
    year = "2014"
}

@article{Arndt:1973ssf,
    author = "Arndt, Richard A. and Long, Dale D. and Roper, L. David",
    title = "{Nucleon-alpha elastic scattering analyses}",
    doi = "10.1016/0375-9474(73)90837-3",
    journal = "Nucl. Phys. A",
    volume = "209",
    pages = "429--446",
    year = "1973"
}

@book{preston1975structure,
  title={Structure of the Nucleus},
  author={Preston, M.A. and Bhaduri, R.K.},
  isbn={9780201059762},
  lccn={lc74013242},
  series={The Advanced Book Program},
  url={https://books.google.de/books?id=ojGBAAAAIAAJ},
  year={1975},
  publisher={Addison-Wesley Publishing Company, Advanced Book Program}
}

\end{document}